\documentclass[twocolumn,showpacs,preprintnumbers,amsmath,amssymb]{revtex4}

\usepackage{graphicx}
\usepackage{dcolumn}
\usepackage{bm}
\usepackage{epsfig}

\def\Mpc{\,{\rm Mpc}}
\def\HO{{100h\,{\rm km\,sec^{-1}\,Mpc^{-1}}}}

\def\la{\mathrel{\mathpalette\fun <}}
\def\ga{\mathrel{\mathpalette\fun >}}
\def\fun#1#2{\lower3.6pt\vbox{\baselineskip0pt\lineskip.9pt
  \ialign{$\mathsurround=0pt#1\hfil##\hfil$\crcr#2\crcr\sim\crcr}}}

\input epsf

\def\plotone#1{\centering \leavevmode
\epsfxsize= 1.0\columnwidth \epsfbox{#1}}

\newenvironment{tablehere}{\def\@captype{table}}{}
\newcommand{\tableskip}{\\[-6pt]}
\def\be{\begin{equation}}
\def\ee{\end{equation}}
\def\ba{\begin{eqnarray}}
\def\ea{\end{eqnarray}}
\def\bea{\begin{eqnarray}}
\def\eea{\end{eqnarray}}

\def\nn{\nonumber}

\def\C{{\cal C}}
\def\cleeu{{\tilde \C}_l^{EE}}
\def\clteu{{\tilde \C}_l^{TE}}
\def\clttu{{\tilde \C}_l^{TT}}

\def\msun{{\,M_\odot}}

\begin{document}

\preprint{}

\title{Dark Energy Tomography}

\author{Yong-Seon Song}
\email{yssong@bubba.ucdavis.edu}
\author{Lloyd Knox}
\email{knox@bubba.ucdavis.edu}
\affiliation{
Department of Physics, One Shields Avenue, University of California, Davis, California 95616
}

\date{\today}

\begin{abstract}
We study how parameter error forecasts for 
tomographic cosmic shear observations are affected by
sky coverage, density of source galaxies, inclusion
of CMB experiments, simultaneou fitting of non--dark energy parameters, 
and the parametrization of the history of the 
dark energy equation-of-state parameter $w(z)$.  
We find tomographic shear-shear power spectra on large angular
scales ($l < 1000$) inferred from all-sky observations,
in combination with Planck, can achieve $\sigma(w_0)=0.06$ and 
$\sigma(w_a) = 0.09$
assuming the equation-of-state parameter is given by 
$w(z) = w_0 + w_a (1-a(z))$
and that nine other matter content and primordial power spectrum
parameters are simultaneously fit.  Taking parameters  
other than $w_0$, $w_a$ and $\Omega_m$ to be completely fixed 
by the CMB we find errors on $w_0$ and $w_a$ that are only 
10\% and 30\% better respectively, 
justifying this common simplifying assumption.  
We also study `dark energy tomography':  reconstuction of $w(z)$ 
assumed to be constant within each of five
independent redshift bins.  With smaller-scale information 
included by 
use of the Jain \& Taylor ratio statistic we find 
$\sigma(w_i) < 0.1$
for all five redshift bins and $\sigma(w_i) <0.02$ 
for both bins at $z < 0.8$.  
Finally, addition of cosmic shear can also reduce errors on
quantities already determined well by the CMB.  We find the sum of 
neutrino masses can be determined to $\pm 0.013$ eV and that
the primordial power specrum power-law index, $n_S$, as well as 
$dn_S/d\ln k$, can be determined more than a factor of two better 
than by Planck alone.  These improvements may be highly valuable
since the lower bound on the sum of neutrino masses is 0.06 eV as
inferred from atmospheric neutrino oscillations, and slow-roll
models of inflation predict non-zero $dn_S/d\ln k$ at the
forecasted error levels when $|n_S-1| > 0.04$.
\end{abstract}



\maketitle

\section{Introduction}

The great mystery of twenty-first century cosmology is the dark
energy--- a smooth, negative-pressure component evoked to bring
the Universe to critical density (e.g.,\cite{dodelson00,bennet03}) and 
to dim high-redshift supernovae \cite{riess98,perlmutter99,perlmutter00}. 
Observers are constraining, or planning to constrain, this curious
component in a number of ways, both through its influence on structure
formation and on geometry.  Here we consider how well the component
can be probed by tomographic cosmic shear observations of large-scale
structure.

Cosmic shear is a very promising technique for studying the large-scale
structure and constraining the cosmological parameters that govern its 
evolution.  For a recent review see \cite{vanwaerbeke03}.  The shear 
signal only depends on the mass distribution; it 
is immune to uncertainties in the relationship between mass and light
that plague other observations.  On large scales, given a model, the
statistical properties of this mass distribution can be calculated with
very high accuracy, thus there is a well-understood relation between theory
and observable.

Cosmic shear also naturally probes the chief redshift range of interest
for constraining dark energy: $0 < z < 2$.  In this redshift range,
dark energy becomes the dominant contributor to the energy density and
thus has significant effect on the expansion rate, $H(z)$.  The expansion
rate affects the shear power spectrum both through its effect on the growth 
rate and also by altering how length scales at a given redshift project
into angular scales.  At $z > 2$, for the simplest models,
the dark energy is highly subdominant and has little influence on 
structure formation.

Multi-band photometry makes possible tomographic cosmic shear, in
which source galaxies are binned according to their
photometrically-determined redshifts, and separate shear maps are made
for each source galaxy redshift bin.  This technique allows one to
study the evolution of the density field as a function of redshift,
restoring the information otherwise lost by projection over the broad
lensing kernel.

Many papers
\cite{hu02a,hu02b,huterer02,refregier03,takada03b,abazajian03,heavens03,benabed03,simon03}
have studied how well tomographic shear correlations can be used to
constrain dark energy parameters.  All but one of these error
forecasts \cite{hu02a} were done in the limit of other cosmological
parameters being perfectly known {\it a priori}.  This assumption
is justified if CMB (or other) observations can be used to determine
the parameters with negligibly small errors.  We investigate 
this assumption by  letting eight other cosmological parameters float, 
constrained only by the cosmic shear data themselves and various
CMB observations.  For WMAP, Planck, and a future
polarization satellite described below we find the constraints on the 
dark energy parameters degrade by 70\%, 30\% and 4\% respectively.  

We parametrize the time-dependence of $w$ in two different ways.  First,
following Linder~\cite{linder02}, we write $w(z) = w_0+ w_a (1-a(z))$.  This 
parametrization has the advantage of tending toward a constant at
high redshift, preventing high-redshift dark energy dominance that
can occur with the parameterization $w(z) = w_0 + w_z z$.  It is 
therefore a convenient parametrization when combining $z < 2$ observations with
$z \simeq 1100$ CMB observations.  

Weller \& Albrecht \cite{weller01} have argued that all models in the
literature can have their luminosity distances as a function of
redshift fit by the $w(z) = w_0 + w_z z$ parametrization to 1\%
accuracy in the $0 < z < 2$ range.  However, dark energy is so poorly
understood that it is worth considering $w(z)$ histories that do not
occur in any existing scalar field quintessence models.  Therefore we
also study a less model-dependent parameterization, one in which
$w(z)$ is constant within each of five independent redshift bins.  An approach
somewhat similar to ours is that of \cite{huterer03} who parametrized
$w(z)$ with its principal components, given supernovae data as
expected from the Supernova/Acceleration Probe (SNAP).  We have found that
data that give similar constraints in the $w_0$, $w_a$ plane may
give strikingly different constraints in the space of redshift bins of $w$.

In this paper we concentrate on what is possible with the data on large
angular scales where theoretical interpretation is most straightforward.
There is also great potential for data on smaller angular scales although
the state of theory must be improved to take full advantage of it.  Potentially
great use can be made of two and three-point functions out to $l=10^5$
(e.g., \cite{refregier03}).  Even at $l < 3000$, including the 3-point
function can improve errors by a factor of 3 \cite{takada03b}.
Counting of mass clusters is another promising way to exploit the small
angular scale data \cite{tyson02}.

We do consider one method of using the small angular scale data here.
Recently \cite{jain03} and \cite{bernstein03} have proposed using the
data in such a way as to be insensitive to the uncertain fluctuation
statistics and only dependent on easily--calculated geometric quantities.  
We combine a conservative use of
the ratio statistic of \cite{jain03} with the large angular scale
shear-shear power spectra.  We find that
the addition of the ratio statistic greatly improves our
reconstruction of $w(z)$ even with a conservative forecast of
how well the ratio statistic can be measured.

In addition to dark energy, another exciting application of
high-precision cosmic shear data is to determine the mass of the relic
neutrinos.  In the first cosmic shear + CMB forecasting papers,
\cite{hu99a,hu99b} it was shown that combining Planck and
all-sky cosmic shear could determine the sum of the masses of the
different flavors of relic neutrinos to 0.02 eV with $w$ fixed to -1.
Recently \cite{kaplinghat03} showed that a post-Planck all-sky CMB
polarization mission could be used to determine the sum to 0.04 eV
without cosmic shear data.  The sum is at least 0.06 eV in order to
explain atmospheric neutrino oscillations \cite{beacom02} hence a
detection is assured, though possibly one of weak statistical
significance.  Here we show that all-sky CMB polarization observations
combined with all-sky weak lensing observations can reduce this error
to 0.027 eV or, with the inclusion of the ratio statistic, to 0.013 eV
even with simultaneous determination of $w_0$ and $w_a$.  This reduction
in error may change a two $\sigma$ detection to a five $\sigma$ detection.

We discuss the shear-shear correlations in section II and
the binning of the dark energy history in section III. In section
IV we discuss the parameters of the future weak lensing and CMB
experiments we consider and in section V outline our error-forecasting
method.  In section VI we give
results for constraints on the dark energy and neutrino mass and all
the other cosmological parameters we assume our simultaneously fit.  
In section VII we see how the results improve with the addition of the ratio
statistic.

\section{tomographic cosmic shear-shear correlations}
The size and shape of galaxies is altered by gravitational
lensing. The effect of lensing is described by the $2\times 2$ 
transformation matrix $A_{ij}$ given by
\ba
A_{ij} \equiv {\partial \theta_s^i \over \partial \theta^j}=
\left(
\begin{array}{cc}
1-\kappa-\gamma_1 & \gamma_2 \\
\gamma_2 & 1-\kappa+\gamma_1 
\end{array}
\right)\,\,.
\ea
where ${\bf \theta}_s$ is the angular location in the source plane
for a light ray appearing at ${\bf \theta}$ in the image plane.
The convergence $\kappa$ describes the magnification or demagnification
while the shear components, $\gamma_1$ and $\gamma_2$, quantify the distortion
of the shape~\cite{kaiser92,bartelmann01}. 

The shear components can be inferred from measurement of galaxy
ellipticities.  Galaxies have their own intrinsic ellipticity and
an additional ellipticity due to lensng so that
\be {\bf e^{\mu}}={\bf e^{\mu}}_{\rm int}+{\bf e^{\mu}}_{\rm lens}
\ee 
where $\mu$ labels the galaxy.
The average shear in a pixel on the sky can be estimated from averaging
over the galaxies in the pixel:
\ba
\gamma_{1}&=&\frac{1}{2N_{\rm pix}}\sum_{\mu}{\rm e^{\mu}_+}
=\frac{1}{2N_{\rm pix}}\sum_{\mu} \big({\rm e^{\mu}_+}_{\rm
,\,int}+{\rm e^{\mu}_+}_{\rm ,\,lens}\big) \nn\\
\gamma_{2}&=&\frac{1}{2N_{\rm pix}}\sum_{\mu}{\rm e^{\mu}_{\times}}
=\frac{1}{2N_{\rm pix}}\sum_{\mu} \big({\rm e^{\mu}_{\times}}_{\rm
,\,int}+{\rm e^{\mu}_{\times}}_{\rm ,\,lens}\big) \ea
where $+$ and $\times$ denote the two orthogonal ellipicity components.
In the absence of correlations between the intrinsic ellipticities, 
the rms error in the measurement of each shear component is
\be 
\sigma(\gamma_1)=\sigma(\gamma_2) =\gamma_{\rm
rms}/\sqrt{N_{\rm pix}} 
\ee 
where $\gamma_{\rm rms}$ is the rms
intrinsic shear of the galaxies and $N_{\rm pix}$ the number of galaxies in
the pixel.  

\begin{figure}[htbp]
\label{fig:shearpower}
  \begin{center}
    \plotone{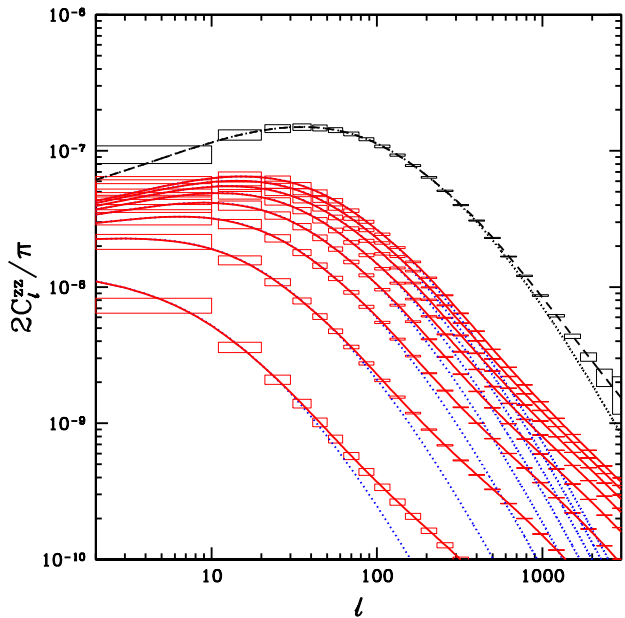}
    \caption{The shear-shear auto power spectra.
The 8 solid curves are the shear power spectra from each of the
galaxy source planes, $B_1$ to $B_8$.  Dotted curves are the linear
perturbation theory approximation.  From bottom to top the source
plane redshift ranges are 
$B_1$: $z\in[0.0,0.4]$, $B_2$: $z\in[0.4,0.8]$,
$B_3$: $z\in[0.8,1.2]$, $B_4$: $z\in[1.2,1.6]$,
$B_5$: $z\in[1.6,2.0]$, $B_6$: $z\in[2.0,2.4]$,
$B_7$: $z\in[2.4,2.8]$ and $B_8$: $z\in[2.8,3.2]$.
The error boxes are forecasts for G4$\pi$ (see Table I).  
The top dashed curve is the shear power spectrum for the CMB source
plane.  The error boxes are forecasts for CMBpol (see Table II).
}
\end{center}
\end{figure}

Maps of the shear components can be decomposed into even parity $E$ modes
and odd parity $B$ modes, just as is the case with linear polarization
Stokes parameters, $Q$ and $U$ \cite{kamionkowski97,seljak97}.  Assuming 
uniform coverage, the noise in 
the $E$ modes will have a diagonal covariance matrix ${\bf N}$ given by 
\be
N_{lmB,l'm'B'}=N_l^{B} \delta_{ll'}\delta_{mm'}\delta_{BB'}.
\ee
where
\be
N_l^{B} = \gamma_{\rm rms}^2 \Omega_{\rm pix}/N_{\rm pix}^B 
\ee
and $B$ labels the source redshift bin.
The signal contribution to the covariance of the shear
$E$ modes, $S$, is correlated across bins, but otherwise diagonal:
\be
S_{lmB,l'm'B'}=C_l^{BB'} \delta_{ll'}\delta_{mm'}.
\ee
The shear angular power spectra are given by
\be
\label{eqn:shearpower}
C_l^{BB'} = 
{\pi^2 l\over 2} \int dr r W^B(r)W^{B'}(r)\Delta_\Phi^2(k,r)
\ee
where the window function for sources in redshift bin, $B$, is
\be
W^B(r) = {1\over \bar{n}_B} {2 \over r} 
\int_{r(z^B_{\rm min})}^{r(z^B_{\rm max})} dr'
{(r'-r)\over r'} N_B(z') dz'/dr'
\Theta(r'-r),
\ee
and the average number density of galaxies in redshift 
bin $B$ is 
\be
\bar{n}_B=\int_{r(z^B_{\rm min})}^{r(z^B_{\rm max})} dr' N_B(z') dz'/dr'.
\ee

We use Peacock and Dodds' formulation~\cite{peacock94} to calculate
the non-linear density power spectrum, $\Delta_\Phi^2(k,r)$.  We only use
data with $l<1000$ to reduce our sensitivity to errors in the non-linear
corrections and departures from Gaussianity \cite{scoccimarro99}.  

We assume that photometrically-determined redshifts can be used
to sort the source galaxies into eight redshift bins with 
$\Delta z=0.4$ from $z=0.0$ to $z=3.2$.  The shear power spectra
for these eight source bins are shown in Fig.~1.  
The higher the source bin redshift, the higher
the shear power.  The shear power spectra are related to the
deflection angle power spectra by 
\be
\label{eqn:deflection}
l(l+1)C_l^{dd}/(2\pi) = 2C_l^{\gamma \gamma}/\pi
\ee
thus the area under the curves in Fig.~1 gives the contribution
to the deflection angle variance from a given range of $l$.
The expected errors on the auto power spectra are given by
\ba
\Delta C_l^{BB}=\sqrt{\frac{2}{(2l+1)f_{sky}}}
\left(C_l^{BB}+N_l^{BB}\right).
\ea

Shear maps can also be inferred from observations of CMB temperature
and polarization.  The top curve in Fig.~1 is the shear power spectrum
for sources at $z\sim 1100$.  The auto and cross-correlation (with
the eight other shear maps) power spectra for the CMB shear map is
also given by Eq.~\ref{eqn:shearpower}.  However, the window
function, due to the small thickness of the last-scattering surface
is simply given by
\ba
W^{\rm CMB}[r(z)] = \frac{r(z_s)-r(z)}{r(z_s)r(z)}\,\,\,
\ea
where $z_s$ is the redshift of the last-scattering surface, defined
as the peak of the visibility function.
We calculate the error in the reconstructed shear maps using the 
quadratic estimator of \cite{hu02}.  
The noise level in Fig.~1 is for the CMBpol experiment described in 
section IV.  At these noise levels, a maximum--likelihood shear map
would have up to two times lower reconstruction noise than the quadratic
estimator shear map \cite{hirata03b}.

Thus we have grouped the data into nine different
redshift bins for lensing tomography:
eight bins with galaxies as the sources (from $z=0.0$ to $z=3.2$) 
and a ninth bin on the CMB last-scattering surface.

\section{binning dark energy history}
We parametrize the history of the dark energy equation of state
parameter as
\ba\label{eq1}
w(z)=\sum_{i} \chi_i(z) \, w_i
\ea
where
\ba
\chi_i(z)= 
 \Big\{\begin{array}{cc}
 1:&\,\,\,\,\,\,\,z^i_{\rm min}<z<z^i_{\rm max}\\ 0:&{\rm otherwise.\,\,\,\,\,\,\,\,\,}
 \end{array}
\ea
We define 5 bins: 4 bins with equal redshift spacing $\Delta z=0.4$ and
the last bin from $z=1.6$ to the last scattering surface.
We choose this binning because the data are sufficiently powerful to
place interesting constraints on this space.  Finer binning, particularly
beyond $z=1.6$, would lead to very strong degeneracies between the bins.

Note that the discrete parametrization of the dark energy equation-of-state
parameter is well-matched with
what we observe in weak lensing experiments with binned source planes.
The variation of $w_i$ in bin $i$ influences the dark energy density in
all the lower redshift bins.  Likewise,  the shear map from galaxies
in bin $i$ are affected by the large-scale 
structure in those same lower redshift bins.

This discrete binning of $w(z)$ does lead to complications
on very large scales where dark energy density fluctuations can be important.
While it is straightforward to calculate the expansion rate
and therefore distance as a function of redshift for $w(z)$ 
given by~Eq.(\ref{eq1}),
it is not clear how to calculate the fluctuations in the dark energy
density and pressure for a model with non-continuous $w(z)$.
On small scales, the dark energy fluctuations (at least for the
simplest scalar--field models) are damped out.
The critical scale $k_Q$ ($l_Q$ in harmonic space)
is calculable from the coupled differential equations of the matter 
and the dark energy fluctuations:
\bea
&&\ddot{\delta}_m+2H\dot{\delta}_m = 4\pi G
(\rho_m\delta_m+\delta\rho_Q+3\delta p_Q), \\
&&\ddot{\delta_Q}+3H\dot{\delta_Q}+
\left(k^2+\frac{d^2 V}{dQ^2}\right)\delta_Q =
\dot{\delta}_m\left[(1+{w}_Q)\rho_Q\right]^{1/2} \nonumber
\eea
where $\delta_m = \delta \rho_m/\rho_m$, $\delta_Q = \delta \rho_Q/\rho_Q$
and a dot denotes derivative with respect to proper time.
In the limit where the inhomogeneity in $Q$-fields is negligible,
the evolution of $\delta_m$ is decoupled from
the dark energy fluctuations~\cite{coble97,caldwell98}.
In this limit, the influence of dark energy is entirely
through $H(z)$.

At $k<\sqrt{|d^2 V/dQ^2|}$, $\delta_Q$ grows via gravitational
instability and will cluster like a matter component. 
The evolution of $\delta_m$ will
be affected by $\delta_Q$ and will therefore be $k$-dependent
in the era of dark energy domination. 
At $k>\sqrt{|d^2 V/dQ^2|}$, $\delta Q$ is damped out
and the evolution of $\delta_m$ will be independent of $\delta Q$.
The characteristic scale $k_Q$ of this regime is given by
\ba
k_Q=0.001\sqrt{(1-{w})
\left(2+2{w}-\frac{{w}\Omega_m}{\Omega_m+\Omega_Qa^{-3{w}}}\right)},
\ea
where $k_Q$ is in the unit of $h\, {\rm Mpc}^{-1}$~\cite{ma99}.

\begin{table}
\begin{center}
\begin{tabular}{c|ccc}
Experiment & f$_{\rm sky}$ & $\bar n_{\rm tot}$ 
& $\bar n_{\rm tot}/\gamma^2_{\rm rms}$ \\
\tableskip\hline\tableskip
G4$\pi$      &1 &65  & 1900 \\
\tableskip\hline\tableskip
G2$\pi$      &0.5 &65 & 1900 \\
\tableskip\hline\tableskip
S300      & 0.0073 &100 & 2920 \\
\tableskip\hline
S1000       &0.024  &100 & 2920 \\
\tableskip\hline
\end{tabular}
\end{center}
\caption{Weak lensing experimental parameters assumed.
'G' and 'S' are for ground- and space-based observations
respectively which have different source redshift distributions.  
Units for the total source sky density, $\bar n_{\rm tot}$, 
are 1/arcmin$^2$ and the per-component rms intrinsic shear, 
$\gamma_{\rm rms}$, is evaluated at $z=1$.}
\end{table}

The characteristic angular wavenumber is given by
$l_Q\sim k_Q r(\eta)$ where $r$ is the angular diameter distance
and $\eta$ is conformal time.
Assuming our fiducial model, $l_Q$ is around 20 for the CMB shear map
and even less for the galaxy source plane shear maps.  We therefore
restrict our analyses to $l > 40$ where the shear maps will be unaffected
by dark energy fluctuations.

Analysis of real data could proceed in two stages.
At the first stage, one would fit the data with the modes $l> 40$
where the effect of $\delta Q$ on matter density perturbations
is negligible.  With the best--fit $w_i$ determined, one could
use interpolation to generate a smooth $w(z)$ as well as the first
few eigenmodes of fluctuations about $w(z)$.  The amplitudes of these
eigenmodes could then be fit to all the data.  

\begin{table}
\begin{center}
\begin{tabular}{ccccccc}
Experiment & $l^{\rm T}_{\rm max}$& $l^{\rm E,B}_{\rm max}$ & $\nu$ (GHz) & $\theta_b$ & $\Delta_T$ & $\Delta_P$\\
\tableskip\hline\tableskip
Planck      &2000 &2500  &  100 & 9.2' & 5.5 & $\infty$ \\
              & &   &  143 & 7.1' & 6  & 11 \\
              & &   &  217 & 5.0' & 13 & 27 \\
\tableskip\hline
CMBpol       &2000  &2500 &  217 & 3.0' & 1  & 1.4 \\
\tableskip\hline
\end{tabular}
\end{center}
\caption{CMB experimental parameters assumed.  Planck has more channels
which we assume are used entirely for control of foregrounds.}
\end{table}

\section{experiments}
Weak lensing surveys are beginning to make their impact on cosmology
\cite{vanwaerbeke03}.  For example, Contaldi et al.\cite{contaldi03} find the
combination of early Red-Sequence Cluster Survey (RCS) data \cite{hoekstra02} and WMAP data
\cite{bennet03} give $\sigma_8=0.89\pm 0.05$ and $\Omega_m=0.30\pm 0.03$.  Ongoing
surveys of tens to hundreds of square degrees include the Deep Lens
Survey (DLS\footnote{The Deep Lens Survey 
\\ {\it http://dls.bell-labs.com/}}), the Canada-France-Hawaii
Telescope Legacy Survey (CFHTLS~\footnote{{\it http://www.cfht.hawaii.edu/Science/CFHLS/}}), 
the National Optical Astronomy Observatory (NOAO\footnote{
{\it
http://www.noao.edu/noao/noaodeep/}}) Deep Wide--Field Survey, 
and the Red-sequence
Cluster Survey (RCS~\footnote{{\it
http://www.astro.utoronto.ca/~gladders/RCS/}}).
Longer term there are projects that may cover 1000 square degrees or more.
These are the SuperNova Acceleration Probe (SNAP~\footnote{
{\it http://snap.lbl.gov/}}),  
Pan-STARRS~\footnote{The Panoramic Survey Telescope $\&$ Rapid Response System 
(28,000 square degrees)\\ {\it http://pan-starrs.ifa.hawaii.edu/}} and
the Large-aperture Synoptic Survey Telescope 
(LSST\footnote{{\it http://www.lsst.org}}).  

We consider two sets of reference surveys, ground-based and space-based
which have different total source density and source redshift distributions. 
For the ground-based surveys we use a galaxy redshift
distribution for a limiting magnitude in R of 26 inferred from 
observations with the Subaru telescope \cite{nagashima02}.  The shape of this
distribution is well-described by the following analytic form: 
\bea
dn/dz &\propto& z^{1.3}\exp\left[-\left(z/1.2\right)^{1.2}\right] \ \ \ {\rm for } \ z <1 \nonumber \\
dn/dz &\propto & z^{1.1}\exp\left[-\left(z/1.2\right)^{1.2}\right] {\rm for } \ z >1.\eea
We use this distribution with the modification that half of the galaxies
in the $1.2 < z < 2.5$ range are discarded for insufficiently accurate 
photometric redshifts.  The amplitude of the distribution
is such that, after this cut, the number density of galaxies 
is 65 per sq. arcmin\footnote{J.A. Tyson, private communication}.  

Space-based surveys have the advantage of no sky noise, allowing
them to reach higher limiting magnitudes, and the possibility of additional
bands in wavelength ranges for which the atmosphere is opaque. The
latter advantage means photometric redshifts can be done even in the
problematic $1.2 < z < 2.5$ range \cite{massey03}.  
For the space-based surveys we assume 
\be
dn/dz \propto z^2 \exp\left(-z/1.5\right)
\ee
which was also assumed by \cite{bernstein03} and 
which is a close approximation to what was used in
a study of the SNAP 300 sq. degree survey \cite{massey03}.
This study predicted a number density of $\sim$100 galaxies per
sq. arcminute. 

The disadvantage of space is that the telescopes are smaller due
to the high cost of putting large mirrors in space.  This means
that ground-based surveys can cover much more sky.  Our two ground-based
surveys are denoted by G4$\pi$ (an all-sky survey) and G2$\pi$ (a half-sky
survey) and our two space-based surveys are called S1000 and S300 with the
numbers denoting the observed area in square degrees. Their properties
are summarized in Table I.  Baseline plans for LSST include a G2$\pi$
survey.

An important quanity for forecasting parameter constraints from {\em any} 
cosmic shear observation is $\gamma_{\rm rms}$.  This quantity is potentially
experiment-dependent, with ground-based observations having a higher effective
$\gamma_{\rm rms}$ than space-based ones \cite{kaiser98,massey03}.  
But for the low-noise ground-based observations we consider here, 
$\gamma_{\rm rms}$ may be just as good as from space.  In fact, ground-based
observations \cite{jarvis03} have achieved very low $\gamma_{\rm rms}$.   
Jarvis et al. \cite{jarvis03} find for the lowest-redshift galaxies in 
their 75 sq. degree CTIO survey that the shape noise contribution 
to the per-component rms shear, $\gamma_{\rm rms}$, is 0.15.  
The shape noise, as inferred from these data, increases approximately
linearly to 0.22 by $z=2$\footnote{D. Wittman, private communication.}.
This trend with redshift is also consistent with simulations of
SNAP weak lensing observations, based on HDF data, that show 
$\gamma_{\rm rms}$, averaged over all galaxies suitable for ellipticity
measurement, increasing from 0.21 to 0.25 with increasing 
observing time per unit solid angle \cite{massey03}.  We therefore
model $\gamma_{\rm rms}$ as
\be
\gamma_{\rm rms}(z) = 0.15+0.035z.
\ee
Note that other forecasting papers typically assume $\gamma_{\rm rms}$
of 0.3 or 0.4.  

For the CMB experiments we consider Planck~\cite{planck} 
and a high-resolution version of 
CMBpol \footnote{CMBpol:
http://spacescience.nasa.gov/missions/concepts.htm.}.
Their specifications are given in Table II.  
We assume that other frequency channels
of Planck and CMBpol (not shown in the table) will clean out non-CMB
sources of radiation perfectly.  Detailed studies have shown foreground
degradation of the results expected from Planck to be mild
\cite{knox99,tegmark00b,bouchet99}.  
At $l>3000$ emission from dusty galaxies will be a significant source
of contamination. The effect is expected to be more severe for
temperature maps. Hence we restrict temperature data to $l<2000$ and
polarization data to $l < 2500$. 

\section{Error Forecasting Method}

The shear two-point functions depend on not just the dark energy
parameters, but on the entire matter content and the primordial power spectrum. 
We do not assume these quantities to be known, but instead assume
that CMB data are available to constrain them.  The CMB power spectra
we include in our analyses are the (unlensed) $\clttu$, $\clteu$ and
$\cleeu$.  We do not use the lensed power spectra to avoid the
complication of the correlation in their errors between different
$\ell$ values and with the error in the CMB-derived shear power
spectrum.  Using the lensed spectra and neglecting these correlations
can lead to overly optimistic forecasts \cite{hu02a}.

To calculate the expected parameter errors we make a
first order Taylor expansion of the parameter dependence of all the
CMB and cosmic shear two--point functions.  
In this `linear response' approximation, given
the expected experimental errors on the power spectra, we can easily
calculate the expected parameter error covariance matrix as the
inverse of the Fisher matrix \cite{jungman96}.
The linear response approximation is improved and susceptibility to
numerical error is reduced with a careful choice of the parameters
used to span a given model space 
\cite{eisenstein99,efstathiou99,hu01a,kosowsky02}.  As in \cite{kaplinghat03}
we take our (non-$w(z)$) set to be 
${\cal P} = \{\omega_m, \omega_b, \omega_\nu,
\theta_s, z_{\rm ri},k^3P_\Phi^i(k_f),n_s,n_s',y_{\rm He}\}$,
with the assumption a flat universe. 
The first three of these are the densities today (in units of
$1.88\times 10^{-29}{\rm g}/{\rm cm}^3$) of cold dark matter plus
baryons, baryons and massive 
neutrinos. We assume two massless species and one massive species.
The next is the angular size subtended by the sound
horizon on the last--scattering 
surface.
The Thompson scattering optical depth for CMB photons, $\tau$, is
parameterized by the redshift of reionzation $z_{\rm ri}$.  
The primordial potential power spectrum is assumed to be
$k^3P_\Phi^i(k) = k_f^3P_\Phi^i(k_f)(k/k_f)^{n_s -1+n_S'ln(k/k_f)}$
with $k_f = 0.05\Mpc^{-1}$.   
The fraction of baryonic mass in Helium 
(which affects the number density of electrons) 
is $y_{\rm He}$.  We Taylor expand about
${\cal P}=\{0.146,0.021,0,0.6,6.3,6.4\times 10^{-11},1,0,0.24\}$.
The Hubble constant for this model is $h=0.655$ where $H_0 = \HO$.

We follow \cite{zaldarriaga97} to calculate the errors expected in
$\clttu$, $\clteu$ and $\cleeu$ given Table 1.
The errors on the unlensed spectra in the regime where lensing is
important (deep in the damping tail) are certainly underestimated
because reconstruction of the unlensed map from the lensed map will
add to the errors. However, this is not worrisome since we limit all
the unlensed spectra to $l < 2000$.

The contribution to the Fisher matrix from the shear-shear correlations
is given by (e.g., \cite{knox01a})
\ba
F^{\gamma\gamma}_{pp'}
 = \sum_{l,B1,B2,B3,B4} {2l+1 \over 2} C_{l,p}^{B1,B2} {\cal W}_l^{B2,B3}
C_{l,p'}^{B3,B4} {\cal W}_l^{B4,B1},
\ea
where the subscript $,p$ denotes differentiation with respect to parameter
$a_p$.  We use ${\cal W}$ to denote thet inverse of the total covariance
matrix:
\be
{\cal W} \equiv ({\bf S} + {\bf N})^{-1}.
\ee 
Note that ${\cal W}$ is an easily invertible block diagonal matrix: 
${\cal W}_{lmB,l'm'B'}={\cal W}_l^{BB'}
\delta_{mm'}\delta_{BB'}$.  The total Fisher matrix
is given by summing $F^{\gamma\gamma}$ with the Fisher matrix for the
unlensed CMB data, $F^{\rm CMB}$.

\section{results}
Above we have described the non-$w(z)$ parameters of our fiducial model, 
about which we consider small fluctuations.  For $w(z)$, we consider
two different fiducial cases.  Both are consant $w$ models, one with  
$w_{\rm fid}=-1$ and the other with $w_{\rm fid}=-0.8$.
About this fiducial model we vary all the cosmological parameters
and either $w_0$ and $w_a$ or the five binned $w_i$'s.

\subsection{Dependence on Sky Coverage and Shape Noise}

First we explore the dependence of constraints on $w_0$ and $w_a$ as
a function of sky coverage and shear weight-per solid angle, 
$\bar n/\gamma_{\rm rms}^2$.  These are both quantities over which the
experimenter has some control.  The shape of $dn/dz$ is taken to be
that assumed for LSST.  The left panels of Fig.~2
show the errors on $w_0$ and $w_a$ assuming $l_{\rm max}=1000$.  The
contours drop rapidly at low $\bar n/\gamma_{\rm rms}^2$ but by 
$\bar n/\gamma_{\rm rms}^2=1000$ the improvement slows dramatically.
At such high shear weights, the dominant contribution to the parameter
errors comes from sample variance rather than shear noise.  

Note that we are only varying the $f_{\rm sky}$ for the cosmic shear
observations.  We keep the $f_{\rm sky}$ for CMB observations fixed
at 1.  Therefore the parameter errors reduce more slowly than 
$f_{\rm sky}^{1/2}$.  The improvement with increasing sky coverage
is slowest for $w_a$ which is more dependent on the constraints provided
by the CMB observations.

If the non-linearities can be controlled out to $l=2000$ the errors on
$w_0$ and $w_a$ improve by about 50\%.  
Because the shear angular power spectrum drops with increasing $l$,
the relative importance of shear noise variance to shear sample variance
increases with increasing $l$.  Therefore the contours in the
right panels of Fig.~2 begin to flatten
out at higher values of $\bar n_{\rm tot}/\gamma_{\rm rms}^2$ than for the 
$l_{\rm max}=1000$ case.

\begin{figure}[htbp]
\label{fig:contours}
  \begin{center}
    \plotone{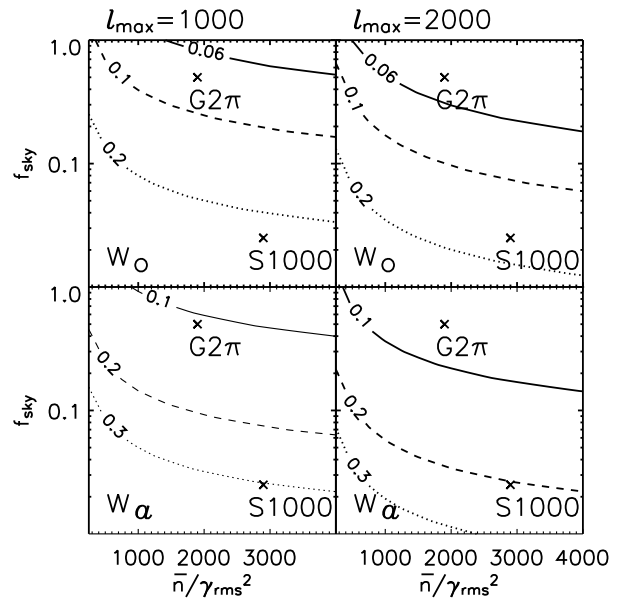}
    \caption{Contours of constant error in $w_0$ (top panels)
and $w_a$ (bottom panels) for cosmic shear observations of
$f_{\rm sky}$ of the sky and shear weight per solid angle
$\bar n_{\rm tot}/\gamma_{\rm rms}^2$.  The left panels are for 
$l_{\rm max}=1000$
and the right panels are for $l_{\rm max}=2000$.  These parameter
error forecasts also include constraints from Planck.
}
\end{center}
\end{figure}

\subsection{Dependence on Parameter Space and CMB Experiment}

Next we turn to forecasted constraints for $w_0$ and $w_a$ for
the four cosmic shear observations combined with the two CMB
observations.  First note from Table III that the CMB experiments 
alone are incapable of constraining these parameters.  The primary
CMB signal offers no constraint due to the angular-diameter distance
degeneracy and CMB lensing only provides a very weak handle since
dark energy is only important at very low redshifts, well below
the peak of the CMB lensing kernel.  The situation improves 
dramatically with the introduction of the cosmic shear-shear 
correlations.

\begin{tablehere}
\begin{table*}[hbt]\small
\caption{\label{tab:lt}}
\begin{center}
\begin{tabular}{c|c|c|c|c|c|c|c|c|c|c|c|c}
\tableskip\hline\hline \tableskip
CMB&Cosmic Shear&
$w_0$ & $w_a$ & $m_\nu$ & $\ln \Omega_m h^2$ & $\ln P_{\Phi}$
&$n_S$&$n_S'$&$\tau$&$\ln \omega_b$&$Y_P$&$\theta_s$\\
\tableskip\hline\tableskip
Planck & none  &  2.2&  3.3&  0.18&  0.0084&  0.017&
  0.0078&  0.0035&  0.010&  0.0086&  0.013&
  0.00016\\
\tableskip\hline
 &  & 0.012 & 0.020 & --- & --- & --- & --- & --- & --- & ---  & --- &
--- \\
none & G4$\pi$ & 0.051 & 0.066 & --- & 0.00089 & --- & --- & --- & --- &
---  & --- & --- \\
 &  & 0.051 & 0.069 & --- & --- & --- & --- & --- & --- & ---  & --- &
0.00027 \\
\tableskip\hline
 & S300  &  0.41&  0.49&  0.12&  0.0068&  0.016&
  0.0066&  0.0034&  0.0091&  0.0076&  0.011&
  0.00014\\
Planck &S1000  &  0.28&  0.34&  0.090&  0.0059&  0.015&
  0.0060&  0.0033&  0.0090&  0.0069&  0.011&
  0.00014\\
&G2$\pi$  &  0.076&  0.11&  0.052&  0.0039&  0.014&
  0.0046&  0.0029&  0.0083&  0.0059&  0.0094&
  0.00012\\
&G4$\pi$  &  0.056&  0.087&  0.045&  0.0035&  0.013&
  0.0044&  0.0025&  0.0077&  0.0058&  0.0087
 &  0.00012\\
\hline\tableskip
 &  & 0.049 & 0.063 & --- & 0.00083 & --- & --- & --- & --- & --- & ---
 & --- \\
 &  & 0.021 & 0.031 & --- & --- & --- & --- & --- & --- & ---  & --- &
 $9.4\times 10^{-5}$\\
 &  & 0.050 & 0.064 & --- & 0.00086 & --- & --- & --- & --- & ---  & ---
&  $9.7\times 10^{-5}$\\
Planck & G4$\pi$ & 0.050 & 0.065 & 0.018 & 0.0015 & --- & --- & --- &
--- & ---  & --- & $9.8\times 10^{-5}$ \\
 & & 0.050 & 0.069 & 0.018 & 0.0028 & 0.0093 & --- & --- & 0.0055 & ---
 & --- & 0.00011 \\
 & & 0.051 & 0.070 & 0.020 & 0.0033 & 0.011 & 0.0020 & --- & 0.0070 &
---  & --- & 0.00011 \\
 & & 0.052 & 0.074 & 0.023 & 0.0033 & 0.011 & 0.0020 & 0.0019 & 0.0070 &
---  & --- & 0.00011 \\
\tableskip\hline
 & none  &  0.79&  1.1&  0.048&  0.0040&  0.017&
  0.0031&  0.0018&  0.0097&  0.0028&  0.0048&
  $4.8\times 10^{-5}$\\
 & S300  &  0.18&  0.24&  0.047&  0.0038&  0.012&
  0.0029&  0.0018&  0.0073&  0.0028&  0.0047&
  $4.6\times 10^{-5}$\\
CMBpol & S1000  &  0.13&  0.17&  0.045&  0.0036&  0.011&
  0.0028&  0.0018&  0.0069&  0.0028&  0.0046&
  $4.6\times 10^{-5}$\\
& G2$\pi$  &  0.064&  0.089&  0.031&  0.0026&  0.0099&
  0.0026&  0.0016&  0.0060&  0.0027&  0.0044&
  $4.2\times 10^{-5}$\\
& G4$\pi$  &  0.049&  0.070&  0.027&  0.0023&  0.0093&
  0.0025&  0.0015&  0.0057&  0.0026&  0.0043&
  $4.1\times 10^{-5}$\\
\tableskip\hline
\end{tabular}\\[12pt]
\begin{minipage}{5.2in}
NOTES.---%
Forecasts of 1-$\sigma$ errors
for the combination of tomographic shear-shear two-point functions with 
Planck and CMBpol.  A `---' indicates a parameter held fixed.  
\end{minipage}
\end{center}
\end{table*}
\end{tablehere}

Unlike luminosity distances to type-Ia supernovae, cosmic shear
power spectra depend on a lot of parameters other than $\Omega_m$,
$w_0$ and $w_a$.  In fact, they depend on all the parameters
in our full set except for the optical depth $\tau$.  Unfortunately,
the spectra are nearly featureless and therefore incapable of
simultaneously fitting this large number of parameters.    
Thus the CMB measurements are key to breaking the degeneracies.

The degeneracy-breaking provided by the CMB is highly effective.
Forecasts for errors on the $w(z)$ parameters, assuming everything
else is fixed by Planck, are fairly accurate as long as they include
one more parameter to be determined by cosmic shear data.  As shown at
the bottom of Table III, this parameter could be $\theta_s$ or
$\Omega_m h^2$.  However, assuming that Planck determines both
$\theta_s$ and $\Omega_m h^2$ perfectly well results in overly
optimistic forecasts for $w_0$ and $w_a$.  This is a bit surprising
since the CMB does such an exquisite job on both $\theta_s$ and
$\Omega_m h^2$.  The reason for it is that the shear power spectra are
highly sensitive to the shape parameter, $\Omega_m h$ which is a
combination of $\theta_s$ and $\Omega_m h^2$.  Even very small
uncertainties in $\Omega_m h$ can significantly increase the errors on
$w_0$ and $w_a$.  
 
Replacing Planck with four-year WMAP, the error forecasts for 
$w_0$ and $w_a$ from G$4\pi$ weaken from 0.056 to 0.064 and from 
0.087 to 0.11 respectively.  Replacing Planck with CMBpol, as we
see in Table III, strengthens the constraint to the limit of all
parameters (other than $w_0$, $w_a$ and either
$\theta_s$ or $\Omega_m h^2$) being fixed by the CMB. 

\subsection{Neutrino Mass}

Replacing a massless neutrino species with a massive one increases the
energy density and therefore the expansion rate, suppressing the
growth of structure.  The suppression of the power spectrum is scale
dependent and the relevant length scale is the Jeans length for
neutrinos \cite{bond83,ma96,hu98} which decreases with time as the
neutrino thermal velocity decreases.  This suppression of growth is
ameliorated at scales larger than the Jeans length at
matter--radiation equality, where the neutrinos can cluster.
Neutrinos never cluster at scales smaller than the Jeans length
today. The net result is no effect on scales larger than the Jeans
length at matter-radiation equality and a scale-indepenent suppression
of power on scales smaller than the Jeans length today.  Both the
total suppression of power and the Jeans length today depend on the
neutrino mass.

The observed galaxy power spectrum (which is
proportional to the matter power spectrum at sufficiently large
scales), combined with CMB observations can be
used to put constraints on $m_\nu$ \cite{hu98a}. At present such an
analysis yields an upper bound on $m_\nu$ of $\sim 0.3$ eV
\cite{spergel03}.

Since the shear power spectra depend on the history of the matter
power spectrum they can also be used to determine neutrino masses.
Recently, \cite{kaplinghat03} showed that an experiment with the 
specifications of CMBpol in Table II could determine the mass of 
a relic neutrino species
via its effect on the lensing of the cosmic microwave background
with an uncertainty of $0.044$eV.  This analysis assumed a 
time-independent $w$.  We see in Table III that allowing for variation
of $w_a$ weakens the neutrino constraint only slightly to 0.048 eV.  
A reasonable prior on $w_0$ and $w_a$ would eliminate this degradation.

Adding constraints from shear-shear correlations to CMBpol only improves
the neutrino mass constraints by less than a factor of 2.  We will
see more dramatic improvement with the introduction of the ratio statistic
in the next section.  Shear-shear correlations make a more dramatic
improvement for $m_\nu$ when added to Planck.  The combination of 
Planck and G$2\pi$ does nearly as well as CMBpol alone.  A similar
result was found in \cite{hu99a,hu99b}, although with the assumption
of $w=-1$.  

\subsection{Primordial Power Spectrum Parameters}

One of the virtues of inflation is the robustness and simplicity
of its predictions. On the other hand, this robustness and
simplicity make it difficult to probe observationally in detail.
One of the few observational handles we have is the shape
of the primordial power spectrum.  High--precision determination
of $n_S$ and (even better) determination of non-zero $n_S' \equiv
dn_S/d\ln k$ would be highly valuable information about inflation
or whatever mechanism generated the initital fluctuations.

Recently the scalar power spectrum has been a subject of
great interest due to claims of non-zero $n_S'$.   The
evidence from CMB data alone is very weak: $n_S' \equiv dn_s/d\ln k =
-0.055 \pm 0.038$ \cite{spergel03}.  Combining CMB data with the Croft
et al. matter power spectrum inferred from high--resolution
observations of the Ly$_\alpha$ forest results in $n_S' = -0.031 \pm
0.017$ \cite{spergel03}.  Other authors \cite{seljak03} working with
the same datasets though have since found looser constraints when 
they marginalize over the
mean ionizing flux as a function of redshift.  The importance of
marginalizing over this parameter, which leads to a large degeneracy
between spectral index and amplitude, was pointed out in
\cite{zaldarriaga01}.

Progress in studying the scalar power spectrum is likely to come
mostly from higher-resolution CMB experiments.  As we
can see in Table III (and in \cite{kaplinghat03}) Planck will improve 
the constraints on $n_S'$
from current values by a factor of 10.  If the error bars shrink
around the current best-fit values, this improved precision will
have profound implications since such large departures from scale-invariance
are not expected in inflationary models.  

The difference $n_S-1$ is first order in the slow-roll parameters and
$n_S'$ is second order.  Thus we expect $|n_S'|$ to be on the order
of $(n_S-1)^2$.  If $1-n_S$ = 0.03, which is perfectly consistent with
present data, then we would expect to see non-zero $n_S'$ at about the
$10^{-3}$ level.  It would be tremendously exciting to actually confirm 
this expectation.  Note though that there are models with 
$|n_S'| \sim n_S-1$ \cite{gold03} so this is not a firm test.  
Nevertheless, finding $\sigma(n_S') >> n_S-1$ would be difficult
to reconcile with inflation \cite{gold03}.

Confirming the expectation that $n_S-1$ is on the order of $(n_S-1)^2$
is likely to take the precision of CMBpol, as pointed out in
\cite{kaplinghat03}.  Unfortunately, even if $1-n_S$ is as large
as 0.06, CMBpol alone would still only achieve a two-sigma detection
of $n_S'=(n_S-1)^2$.  Further, the addition of cosmic shear to CMBpol does
not improve the constraints on $n_S'$ by much, although it can
improve $\sigma(n_S')$ by a factor of five over what's
possible with WMAP data alone \cite{ishak03}.

It is very hard to imagine what could improve the determination
of $n_S'$ further \cite{knox03}.  Extending dynamic range further only improves
the error logarthmically, and then only if power at the smaller
scale is measured with precision comparable to that at the larger scale.
It is highly unlikely that systematics on any small-scale measurement
could be controlled well enough to result in constraints on the
primordial power spectrum amplitude at the $10^{-3}$ level.

Finally, note that the amplitude of the primordial power spectrum is
better-determined with the addition of cosmic shear.  And since the 
CMB determines $P_\Phi \exp(-2\tau)$ to very high accuracy, the $\tau$
constraint improves too \cite{hu02a}.

\subsection{Dark Energy Tomography}

Now we turn to the results for the discrete parametrization of $w(z)$,
shown in Fig.~3 and Fig.~4.
We see that the errors grow with increasing redshift as expected due to the
decreasing importance of dark energy with increasing redshift.
For $w=-1$, matter-dark energy equality is at $z_{eq}\simeq 0.25$.
The percentages of the total density in dark energy
are $25\%$, $7\%$ and $3\%$ at $z=1$, $z=2$ and $z=3$ respectively.
For $w=-0.8$ the drop off is slower so the rise in the errors is
slower.  The highest-redshift bin is the broadest bin, stretching
from $z=1.6$ all the way to $z=1100$; this width is responsible for
the smaller error in the highest redshift bin.   

\begin{figure}[htbp]
\label{fig:wbin1}
  \begin{center}
    \plotone{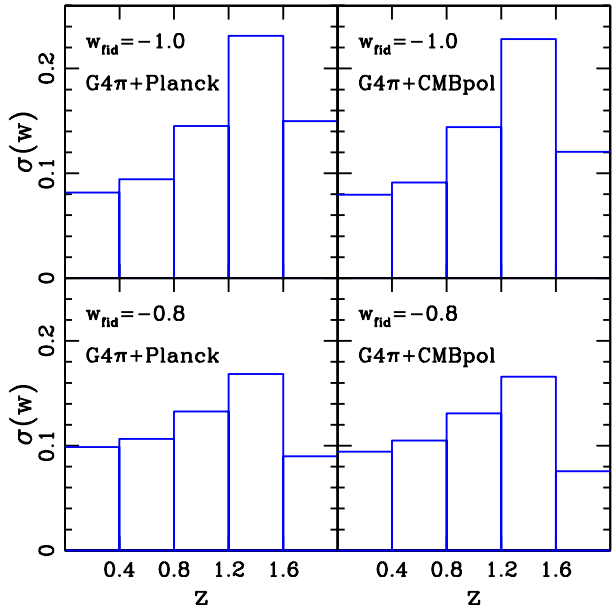}
    \caption{Error boxes indicate expected 1-$\sigma$ error in 
each $w_i$ for G$4\pi$ combined with either Planck (left panels)
or CMBpol (right panels) for $w_{\rm fid}=-1$ (top panels) or
-0.8 (bottom panels).  
}
\end{center}
\end{figure}

\begin{figure}[htbp]
\label{fig:wbin2}
  \begin{center}
    \plotone{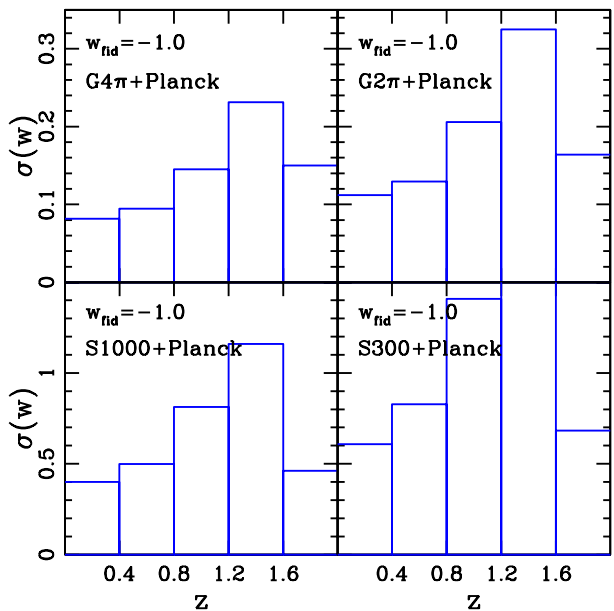}
    \caption{Error boxes indicate expected 1-$\sigma$ error in 
each $w_i$ for the combination of experiments specified in each
panel.}
\end{center}
\end{figure}

We see from Fig.~3 and Fig.~4 that
weak lensing observations of very large amounts of sky will
allow for a reconstruction of $w(z)$ up to $z=2$, with errors
in all five bins (for the G4$\pi$, $w_{\rm fid}=-0.8$ case) less
than 0.2.  

The smaller sky coverage of the space-based surveys leads to much
worse constraints.  Although here we must remind the reader that we
are focusing on science that can be done with information on the
larger angular scales.  Space-based observations will have 
advantages with respect to ground--based observations at smaller
angular scales.

The errors in $w_i$ are highly correlated.
It is useful for us to find an uncorrelated linear combination of $w_i$
when we fit the large scale modes at the second stage.
We write $\tilde w_{\alpha}=R_{\alpha i}w_i$
and find $R_{\alpha i}$ such that
\ba
\langle\Delta\tilde w_{\alpha}\Delta\tilde w_{\beta}\rangle=
\tilde F_{\alpha\beta}^{-1}&=&E^2_{\alpha}\delta_{\alpha\beta}\nn\\
&=& \sum_{i,j=1}^5 R_{\alpha i}F_{ij}^{-1}R_{j\beta}
\ea
where $\tilde F_{\alpha\beta}^{-1}$ is the (diagonal) Fisher matrix for
the transformed parameters and 
the transformation matrix $R$ satisfies 
$R_{\alpha i}R^T_{i \beta}=\delta_{\alpha\beta}$.
TABLE~IV shows 1-$\sigma$ errors for the amplitudes of the
eigenmodes of $w_i$.

\begin{table}
\label{tab:eigenvalues}
\begin{center}
\begin{tabular}{c|c|c|c|c}
\tableskip\hline\hline \tableskip   E1 &E2 &E3&E4&E5 \\
\tableskip\hline\tableskip
0.021&  0.046&  0.088& 0.13& 0.28 \\
\tableskip\hline
\end{tabular}
\end{center}
\caption{Forecasts for 1-$\sigma$ errors on the $w_i$ eigenmode amplitudes 
for G$4\pi$ and CMBpol assuming $w_{\rm fid}=-1$.}
\end{table}

After the first stage of analysis, restricted to $l > 40$ and ignoring
dark energy fluctuations, one would proceed to a second stage which would
extend to larger scales and take the fluctuations into account.  The
second stage of analysis would use a smoothed version of the eigenmodes 
derived from the first stage of analysis as the basis for the parametrization 
of $w(z)$.  We can test the consistency from what we found at the
first stage, though we do not expect noticeable improvement at the
second stage.  Because the large-scale and small-scale data are uncorrelated,
the small-scale data could simply be included as a prior determined
in the first stage of the analysis.  Icluding the small-scale data
in this way may be of practical use, in terms of making the data analysis
tractable, if the parameter space needs to be broadened to include
the sound-speed of the dark energy.  For a study of how well the sound
speed can be constrained see \cite{hu02a}.


\section{the ratio statistic}

We have restricted our use of the shear maps to $l < 1000$ to
reduce our sensitivity to theoretical errors in the calculation
of $P(k,z)$.  The density power spectrum is difficult to calculate
on small scales where the influence of baryons becomes important.

However, discarding the data at $l > 1000$ is highly unfortunate since
there is tremendous signal-to-noise on these angular scales, as
can be seen in Fig.~1.  Recently Jain \& Taylor~\cite{jain03} proposed a ratio 
statistic that uses this small-scale data to constrain
cosmology, but that is not biased by errors in the small-scale power 
spectrum.  Here we combine the ratio statistic with the shear-shear
power spectra at $l < 1000$ and find it provides highly complementary
constraints on the bins of $w$.  

\begin{figure}[htbp]
\label{fig:wi-contours}
  \begin{center}
    \plotone{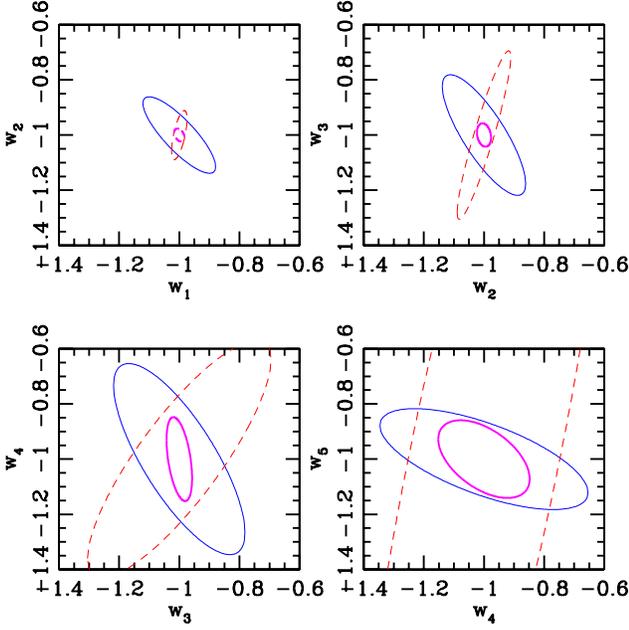}
    \caption{Forecasts of 1-$\sigma$ error contours for pairs of $w_i$
assuming G$4\pi$ and CMBpol.
The thin solid contour is for shear-shear correlations alone,
the dashed contour is for the ratio statistic alone and
the thick contour is for their combination.
}
\end{center}
\end{figure}

Further details on how the Jain \& Taylor idea~\cite{jain03} might be
implemented are given in Bernstein \& Jain~\cite{bernstein03}.  They
write the observed eccentricity as
\be
{\bf e}_k^{\rm obs}=\sum_l d_l({\bf x}_k)g_{lk} + {\bf e}_k.
\ee
The distortion field $d_l$ is the
distortion field for sources at infinity due to mass inside redshift
shell $l$.  The $g_{lk}$ is a purely geometric term that corrects the
distortion for a source at finite distance $r_k$.  If the distortion
field is dominated by a single lensing object then we can write
\be
{\bf e}_k^{\rm obs} = d_{l*}({\bf x}_k)g_{l*k}+\sum_{l \ne l*} d_{l}({\bf x}_k)g_{lk} + {\bf e}_k
\ee
which is just rewriting the above, but with the understanding that the
term with the sum is now small compared to the others.  With
the assumption of a shape for the distortion field, $d_{l*}({\bf x}_k)$,
one can determine $g_{l*k}$ up to a normalizing constant, as well as
$g_{l*k'}$ up to the same normalizing constant. 
Thus Jain and Taylor proposed using
the ratio of such determinations to give a purely geometric 
quantity.  This is their ratio statistic.  Its expectaion value is
\be
R^l_{kk'} = g_{lk}/g_{lk'}.
\ee

A rough idea of how well $R^l_{kk'}$ can be determined is given by considering
the typical amplitude of the shear signal and the area over which it
can be measured, the latter affecting how many galaxies can be used
to beat down shape noise.  This leads to 
\ba
C^{l}_{kk',jj'}&\equiv&
\Bigg{\langle}\frac{\Delta R^l_{kk'}}{R^l_{kk'}}
\frac{\Delta R^{l}_{jj'}}{R^{l}_{jj'}}\Bigg{\rangle}\nn\\
&\simeq&\frac{\langle n^2_{k}\rangle}
{\gamma^l_{k}\gamma^l_{j}}\delta_{kj}
+\frac{\langle n^2_{k'}\rangle}
{\gamma^l_{k'}\gamma^l_{j'}}\delta_{k'j'}
\ea
where $\gamma^l_{k}$ is the average amplitude of the tangential shear 
of galaxies in source bin $k$ due to lenses in lens slice $l$.   
The noise variance $\langle n^2_{k}\rangle$ is given by
\ba
\langle n^2_{k}\rangle\simeq
\frac{\gamma_{\rm rms}^2}{{\cal N}_k f_l}
\ea
where $f_l$ denotes the fraction of the sky covered
by templates for lenses in lens redshift slice $l$. 

To proceed further we need estimates for the shear template amplitudes
$\gamma_k^l$ and the fraction of the sky in the observed regions
covered by the templates, $f_l$.  Jain \& Taylor \cite{jain03} point
out that the number density of clusters with mass in a logarithmic
intermal about $m=10^{14}\msun$ out to $z=1$ is 0.01/arcmin$^2$.  
Further, for a
galaxy cluster of this mass at $z=0.2$ with lens source at $z=1$
the tangential shear signal is nearly constant out to $1'$ with an 
area-weighted average amplitude of about 0.05 \cite{takada03a}.  
We therefore set $\gamma_k^l=0.05$
and define the area useable for estimating the template amplitude to
be that within the 1' radius.
With this template area per object and considering objects with
masses in the decade surrounding $m= 10^{14}\msun$ we get a fraction
of the observing area covered with templates of about 0.06.  Dividing
this into three lens slices (centered on $z=0.2$, $z=0.6$ and $z=1$) 
we get $f_l = 0.02$.  

Jain and Taylor\cite{jain03} used a full halo model calculation of
the signals expected and area that could be used for measuring
them in each lens slice for each mass of cluster in order to
forecast errors on their ratio statistic, and therefore on dark energy
parameters.  They also used a version of the simple forecast above
that was able to reproduce the results of their more detailed calculation.
Our simple forecast for errors on the ratio statistic is much more conservative
than theirs.  The main difference is that they assumed an area with
a radius of 3' about each cluster could be used for measuring the
shear template amplitude.  We chose
the smaller radius to reduce contamination from projection.  

Our results including lensing cosmography are more speculative than
our results from just CMB and the shear two-point functions.  Projection
effects may be important, even with our restriction to 1'.  Although
\cite{bernstein03} included projection effects, they did so with a 
halo model calculation which may be inadequate for this purpose 
given its assumption of spherical halos.  
Further, the construction of templates from the galaxy distribution may not
be as robust as assumed in \cite{bernstein03}.  Numerical simulations
could address the projection problem \cite{tyson02}.  
Finally, unlike the sample-variance dominated shear-shear correlation 
statistics, the errors we forecast on the ratio statistic depend strongly 
on the achievement of the low level of shape noise that we assume.

\begin{tablehere}
\begin{table*}[hbt]\small
\caption{\label{tab:lt+lc}}
\begin{center}
\begin{tabular}{c|c|c|c|c|c|c|c|c|c|c|c|c}
\tableskip\hline\hline \tableskip
CMB&Cosmic Shear&
$w_0$ & $w_a$ & $m_\nu$ & $\ln \Omega_m h^2$ & $\ln P_{\Phi}$
&$n_S$&$n_S'$&$\tau$&$\ln \omega_b$&$Y_P$&$\theta_s$\\
\tableskip\hline\tableskip
 & none  &  2.2&  3.3&  0.18&  0.0084&  0.017&
 0.0078&  0.0035&  0.010&  0.0086&  0.013&
  0.00016\\
 & S300  &  0.16&  0.26&  0.074&  0.0061&  0.015&
  0.0065&  0.0033&  0.0090&  0.0073&  0.011&
  0.00014\\
Planck &S1000  &  0.10&  0.16&  0.051&  0.0048&  0.015&
  0.0058&  0.0031&  0.0089&  0.0067&  0.011&
  0.00014\\
&G2$\pi$  &  0.031&  0.053&  0.019&  0.0033&  0.014&
  0.0037&  0.0020&  0.0080&  0.0058&  0.0067&
  0.00012\\
&G4$\pi$  &  0.022&  0.040&  0.014&  0.0030&  0.013&
  0.0033&  0.0017&  0.0074&  0.0056&  0.0059&
  0.00011\\
\tableskip\hline
 & none  &  0.79&  1.1&  0.048&  0.0040&  0.017&
  0.0031&  0.0018&  0.0097&  0.0028&  0.0048&
  $4.8\times 10^{-5}$\\
 & S300  &  0.12&  0.20&  0.044&  0.0036&  0.012&
  0.0029&  0.0018&  0.0073&  0.0028&  0.0046
 &  $4.6\times 10^{-5}$\\
CMBpol & S1000  &  0.082&  0.13&  0.037&  0.0033&  0.011&
  0.0028&  0.0018&  0.0066&  0.0028&  0.0046&
  $4.5\times 10^{-5}$\\
& G2$\pi$  &  0.028&  0.047&  0.017&  0.0022&  0.0086&
  0.0024&  0.0014&  0.0053&  0.0026&  0.0037
 &  $3.9\times 10^{-5}$\\
& G4$\pi$  &  0.021&  0.035&  0.013&  0.0021&  0.0080&
  0.0021&  0.0013&  0.0049&  0.0025&  0.0033&
  $3.6\times 10^{-5}$\\
\tableskip\hline
\end{tabular}\\[12pt]
\begin{minipage}{5.2in}
NOTES.---%
Forecasts of 1-$\sigma$ errors
for the combination of tomographic shear-shear two-point functions and the
ratio statistic with Planck (upper table) and CMBpol (lower table).
\end{minipage}
\end{center}
\end{table*}
\end{tablehere}

\begin{figure}[htbp]
\label{fig:lt+lc}
  \begin{center}
    \plotone{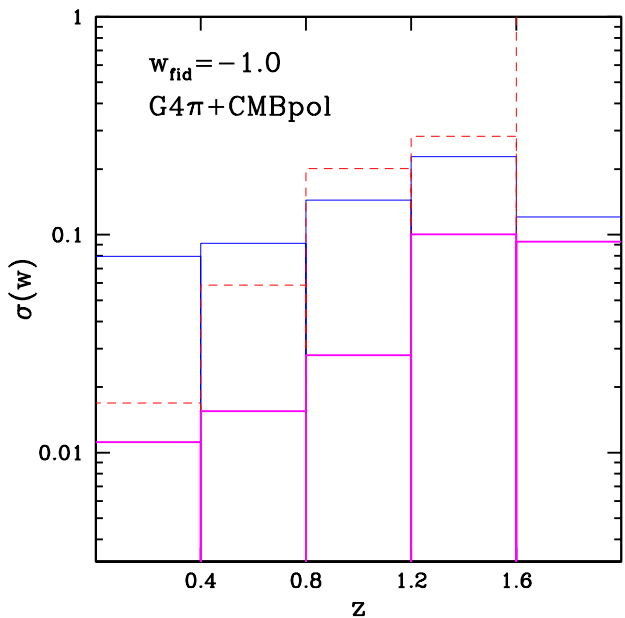}
    \caption{Forecasts of 1-$\sigma$ errors 
in $w_i$ for the $i=1$ to 5 redshift bins for CMBpol combined with
the all-sky weak lensing experiment G$4\pi$.
The thin solid error boxes are for shear-shear correlations alone,
the dash error boxes are for the ratio statistic alone and
the thick solid boxes are for the combination.
}
\end{center}
\end{figure}

The Fisher matrix for the ratio statistic is given by
\ba
F^{\rm R}_{pp'}=\sum_l \sum_{kk'}
\frac{R^l_{kk',p}}{R^l_{kk'}}
C^{l\,\,-1}_{kk',kk'}
\frac{R^l_{kk',p'}}{R^l_{kk'}}
\ea
where we have approximated the $C$ matrix to be independent of cosmological
parameters.  We calculate this Fisher matrix summing over the three lens
planes and all eight cosmic shear source planes.  Because the
ratio statistic is influenced by data on smaller angular scales than
the shear-shear statistics (since there we restricted to $l < 1000$)
the errors are independent and the Fisher matrices simply add:
\ba
F_{pp'}=F^{\gamma\gamma}_{pp'}+F^{\rm R}_{pp'}+F^{\rm CMB}_{pp'}.
\ea

Despite our more conservative treatment, we still find the ratio
statistic to provide a powerful constraint on dark energy especially
when combined with the shear-shear correlations investigated above.
Parameter error forecasts for the two-parameter dark energy model
from combining shear-shear correlations
and the ratio statistic with CMB data 
are shown in Table V.  We see that the addition of the ratio 
statistic reduces errors in the dark energy parameters by about 
a factor of two.  Constraints on neutrino mass also improve
by factors of two and three.  

Even more dramatic improvements occur in the $w_i$ space.
The combination of shear-shear correlations and the ratio
statistic dramatically reduces the degeneracies
of these parameters.
Fig.~5 shows that with shear-shear correlations
the errors in $w_i$ and $w_{i+1}$ are anti-correlated
while with the ratio statistic the errors are positively correlated.
The result for the combination is
a large reduction of the error for each $w_i$ as can be seen in 
Fig.~6.
As is shown in~Fig.~6, G4$\pi$ plus CMBpol
can reconstruct $w(z)$ with uncertainty $\sigma(w_i)\la 0.1$ 
up to $z\sim 2$.

The anti-correlation 
from shear-shear correlations is expected since a positive 
variation of $w_i$
can be compensated for by a negative variation of $w_{i+1}$.
But the complicated ratio statistic leads to
the unusual postive correlation, where a positive variation of
$w_i$ can be partially compensated for by a positive variation in $w_{i+1}$.

\section{Discussion and Conclusions}

The statistical properties of the large-scale density field are sensitive 
to the dark energy equation of state.  Tomographic cosmic shear observations
on large angular scales offer a theoretically robust means of taking
advantage of this dependence.  While the statistical
properties are also sensitive to other matter components and the
primordial power spectra, in combination with CMB observations dark
energy parameters can still be precisely measured despite these other
degrees of freedom.  

We have seen that very tight constraints can be achieved on
the parameters of an 11-dimensional space, including the
equation-of-state parameter and its time variation, the mass
of the relic neutrinos, and the shape and amplitude of the primordial
power spectrum.  Cosmic shear surveys are critical to the
dark energy constraints, and constribute significantly to
the other constraints, especially those on neutrino mass.  

To explore the constraints possible on the history of the 
dark energy equation-of-state parameter we introduced a discrete
parametrization of $w(z)$.  The combination of all-sky weak lensing
observations on large angular scales and all-sky CMB observations
can constrain $w$ in each of five bands (four evenly spaced from
0 to 1.6 and one from 1.6 to last-scattering) to better than 0.2.  

Such precision on $w$ out to redshifts where the dark energy density
is subdominant may prove to be highly illuminating as to the nature
of the dark energy.  Some attractor solutions predict a sharp 
transition of $w$ from $w\ga 0$ to $w=-1$
at a redshift where the dark energy density is still subdominant
\cite{albrecht00,armendariz00,farrar03}.
If such a transition happens at $z<2$, 
then we can probe that transition point with the binned $w_i$.

While we have focused on what is possible with the statistical
properties that can be most robustly predicted from theory, there
is much signal to noise at smaller angular scales.  Further work
is needed to make full use of these data where non-linear corrections
are large.  We studied one means of including information from
small-scales, the ratio statistic of \cite{jain03}, and found it
resulted in large reductions in forecasted parameter errors.  The
$m_\nu$ constraint for G$4\pi$+Planck improved by a factor of three
to the very interesting level of 0.014 eV.  The G$4\pi$ + CMBpol
constraints on the $w_i$ improved even more dramatically in the
three lowest redshift bins.

Many challenges remain to the achievement of the parameter errors
we forecast here.  We have assumed that residuals from
fitting anisotropic point-spread functions, distortions introduced by
the camera, photometric redshift errors, redshift-dependent
calibration errors \cite{hirata03c} and intrinsic alignments
\cite{heavens00,catelan01,crittenden01,pen00,mackey02,jing02} are
negligible. The validity of these assumptions depends on the skill of
the experimenters and analysts and to some degree on the kindness of
nature.  Study of the impact of these non-idealities and
how to tackle them is on-going.  Calibration errors of 5\%, achievable with current data analysis methods
\cite{hirata03c}, lead to only small changes in forecasted
parameter errors \cite{ishak03}.  Intrinsic alignment contamination
can be avoided by using photometrically-determined redshifts to 
exclude spatially-close galaxy pairs\footnote{J.A. Tyson, private
communication} or, more radically, by restricting analyis to shear
cross-spectra and ignoring the auto spectra \cite{takada03c}.  
Avoiding significant biases in photometric redshifts will require
calibration with large, but achievable, numbers of spectroscopic 
redshifts \cite{bernstein03}.

\section{Acknowledgments}
We thank Neal Dalal, Joseph Hennawi, Wayne Hu, Bhuvnesh Jain, Manoj
Kaplinghat, Lori Lubin, Albert Stebbins, Masahiro Takada, J. Anthony
Tyson and David Wittman for useful discussions.  This work was supported by
NASA grant NAG5-11098 and the NSF.

\bibliography{/work6/yssong/Paper/bib/mybib}

\begin{thebibliography}{75}
\expandafter\ifx\csname natexlab\endcsname\relax\def\natexlab#1{#1}\fi
\expandafter\ifx\csname bibnamefont\endcsname\relax
  \def\bibnamefont#1{#1}\fi
\expandafter\ifx\csname bibfnamefont\endcsname\relax
  \def\bibfnamefont#1{#1}\fi
\expandafter\ifx\csname citenamefont\endcsname\relax
  \def\citenamefont#1{#1}\fi
\expandafter\ifx\csname url\endcsname\relax
  \def\url#1{\texttt{#1}}\fi
\expandafter\ifx\csname urlprefix\endcsname\relax\def\urlprefix{URL }\fi
\providecommand{\bibinfo}[2]{#2}
\providecommand{\eprint}[2][]{\url{#2}}

\bibitem[{\citenamefont{{Dodelson} and {Knox}}(2000)}]{dodelson00}
\bibinfo{author}{\bibfnamefont{S.}~\bibnamefont{{Dodelson}}} \bibnamefont{and}
  \bibinfo{author}{\bibfnamefont{L.}~\bibnamefont{{Knox}}},
  \bibinfo{journal}{Physical Review Letters} \textbf{\bibinfo{volume}{84}},
  \bibinfo{pages}{3523} (\bibinfo{year}{2000}).

\bibitem[{\citenamefont{{Bennett} et~al.}(2003)\citenamefont{{Bennett},
  {Halpern}, {Hinshaw}, {Jarosik}, {Kogut}, {Limon}, {Meyer}, {Page},
  {Spergel}, {Tucker} et~al.}}]{bennet03}
\bibinfo{author}{\bibfnamefont{C.~L.} \bibnamefont{{Bennett}}},
  \bibinfo{author}{\bibfnamefont{M.}~\bibnamefont{{Halpern}}},
  \bibinfo{author}{\bibfnamefont{G.}~\bibnamefont{{Hinshaw}}},
  \bibinfo{author}{\bibfnamefont{N.}~\bibnamefont{{Jarosik}}},
  \bibinfo{author}{\bibfnamefont{A.}~\bibnamefont{{Kogut}}},
  \bibinfo{author}{\bibfnamefont{M.}~\bibnamefont{{Limon}}},
  \bibinfo{author}{\bibfnamefont{S.~S.} \bibnamefont{{Meyer}}},
  \bibinfo{author}{\bibfnamefont{L.}~\bibnamefont{{Page}}},
  \bibinfo{author}{\bibfnamefont{D.~N.} \bibnamefont{{Spergel}}},
  \bibinfo{author}{\bibfnamefont{G.~S.} \bibnamefont{{Tucker}}},
  \bibnamefont{et~al.} (\bibinfo{year}{2003}),
  \bibinfo{note}{astro-ph/0302207}.

\bibitem[{\citenamefont{{Riess} et~al.}(1998)\citenamefont{{Riess},
  {Filippenko}, {Challis}, {Clocchiatti}, {Diercks}, {Garnavich}, {Gilliland},
  {Hogan}, {Jha}, {Kirshner} et~al.}}]{riess98}
\bibinfo{author}{\bibfnamefont{A.~G.} \bibnamefont{{Riess}}},
  \bibinfo{author}{\bibfnamefont{A.~V.} \bibnamefont{{Filippenko}}},
  \bibinfo{author}{\bibfnamefont{P.}~\bibnamefont{{Challis}}},
  \bibinfo{author}{\bibfnamefont{A.}~\bibnamefont{{Clocchiatti}}},
  \bibinfo{author}{\bibfnamefont{A.}~\bibnamefont{{Diercks}}},
  \bibinfo{author}{\bibfnamefont{P.~M.} \bibnamefont{{Garnavich}}},
  \bibinfo{author}{\bibfnamefont{R.~L.} \bibnamefont{{Gilliland}}},
  \bibinfo{author}{\bibfnamefont{C.~J.} \bibnamefont{{Hogan}}},
  \bibinfo{author}{\bibfnamefont{S.}~\bibnamefont{{Jha}}},
  \bibinfo{author}{\bibfnamefont{R.~P.} \bibnamefont{{Kirshner}}},
  \bibnamefont{et~al.}, \bibinfo{journal}{Astron. J.}
  \textbf{\bibinfo{volume}{116}}, \bibinfo{pages}{1009} (\bibinfo{year}{1998}).

\bibitem[{\citenamefont{{Perlmutter} et~al.}(1999)\citenamefont{{Perlmutter},
  {Turner}, and {White}}}]{perlmutter99}
\bibinfo{author}{\bibfnamefont{S.}~\bibnamefont{{Perlmutter}}},
  \bibinfo{author}{\bibfnamefont{M.~S.} \bibnamefont{{Turner}}},
  \bibnamefont{and} \bibinfo{author}{\bibfnamefont{M.}~\bibnamefont{{White}}},
  \bibinfo{journal}{Physical Review Letters} \textbf{\bibinfo{volume}{83}},
  \bibinfo{pages}{670} (\bibinfo{year}{1999}).

\bibitem[{\citenamefont{{Perlmutter} and {SNAP
  Collaboration}}(2000)}]{perlmutter00}
\bibinfo{author}{\bibfnamefont{S.}~\bibnamefont{{Perlmutter}}}
  \bibnamefont{and} \bibinfo{author}{\bibnamefont{{SNAP Collaboration}}},
  \bibinfo{journal}{Bulletin of the American Astronomical Society}
  \textbf{\bibinfo{volume}{32}}, \bibinfo{pages}{1504} (\bibinfo{year}{2000}).

\bibitem[{\citenamefont{{Van Waerbeke} and {Mellier}}(2003)}]{vanwaerbeke03}
\bibinfo{author}{\bibfnamefont{L.}~\bibnamefont{{Van Waerbeke}}}
  \bibnamefont{and}
  \bibinfo{author}{\bibfnamefont{Y.}~\bibnamefont{{Mellier}}},
  \bibinfo{journal}{ArXiv Astrophysics e-prints}  (\bibinfo{year}{2003}),
  \eprint{astro-ph/0305089}.

\bibitem[{\citenamefont{{Hu}}(2002)}]{hu02a}
\bibinfo{author}{\bibfnamefont{W.}~\bibnamefont{{Hu}}}, \bibinfo{journal}{\prd}
  \textbf{\bibinfo{volume}{65}}, \bibinfo{pages}{23003} (\bibinfo{year}{2002}).

\bibitem[{\citenamefont{{Hu} and {Okamoto}}(2002{\natexlab{a}})}]{hu02b}
\bibinfo{author}{\bibfnamefont{W.}~\bibnamefont{{Hu}}} \bibnamefont{and}
  \bibinfo{author}{\bibfnamefont{T.}~\bibnamefont{{Okamoto}}},
  \bibinfo{journal}{\apj} \textbf{\bibinfo{volume}{574}}, \bibinfo{pages}{566}
  (\bibinfo{year}{2002}{\natexlab{a}}).

\bibitem[{\citenamefont{{Huterer}}(2002)}]{huterer02}
\bibinfo{author}{\bibfnamefont{D.}~\bibnamefont{{Huterer}}},
  \bibinfo{journal}{\prd} \textbf{\bibinfo{volume}{65}}, \bibinfo{pages}{63001}
  (\bibinfo{year}{2002}).

\bibitem[{\citenamefont{{Refregier} et~al.}(2003)\citenamefont{{Refregier},
  {Massey}, {Rhodes}, {Ellis}, {Albert}, {Bacon}, {Bernstein}, {McKay}, and
  {Perlmutter}}}]{refregier03}
\bibinfo{author}{\bibfnamefont{A.}~\bibnamefont{{Refregier}}},
  \bibinfo{author}{\bibfnamefont{R.}~\bibnamefont{{Massey}}},
  \bibinfo{author}{\bibfnamefont{J.}~\bibnamefont{{Rhodes}}},
  \bibinfo{author}{\bibfnamefont{R.}~\bibnamefont{{Ellis}}},
  \bibinfo{author}{\bibfnamefont{J.}~\bibnamefont{{Albert}}},
  \bibinfo{author}{\bibfnamefont{D.}~\bibnamefont{{Bacon}}},
  \bibinfo{author}{\bibfnamefont{G.}~\bibnamefont{{Bernstein}}},
  \bibinfo{author}{\bibfnamefont{T.}~\bibnamefont{{McKay}}}, \bibnamefont{and}
  \bibinfo{author}{\bibfnamefont{S.}~\bibnamefont{{Perlmutter}}},
  \bibinfo{journal}{ArXiv Astrophysics e-prints}  (\bibinfo{year}{2003}),
  \eprint{astro-ph/0304419}.

\bibitem[{\citenamefont{{Takada} and {Jain}}(2003{\natexlab{a}})}]{takada03b}
\bibinfo{author}{\bibfnamefont{M.}~\bibnamefont{{Takada}}} \bibnamefont{and}
  \bibinfo{author}{\bibfnamefont{B.}~\bibnamefont{{Jain}}},
  \bibinfo{journal}{ArXiv Astrophysics e-prints}
  (\bibinfo{year}{2003}{\natexlab{a}}), \eprint{astro-ph/0304034}.

\bibitem[{\citenamefont{{Abazajian} and {Dodelson}}(2003)}]{abazajian03}
\bibinfo{author}{\bibfnamefont{K.}~\bibnamefont{{Abazajian}}} \bibnamefont{and}
  \bibinfo{author}{\bibfnamefont{S.}~\bibnamefont{{Dodelson}}},
  \bibinfo{journal}{Physical Review Letters} \textbf{\bibinfo{volume}{91}},
  \bibinfo{pages}{41301} (\bibinfo{year}{2003}).

\bibitem[{\citenamefont{{Heavens}}(2003)}]{heavens03}
\bibinfo{author}{\bibfnamefont{A.}~\bibnamefont{{Heavens}}},
  \bibinfo{journal}{ArXiv Astrophysics e-prints} pp. \bibinfo{pages}{4151--+}
  (\bibinfo{year}{2003}).

\bibitem[{\citenamefont{{Benabed} and {Van Waerbeke}}(2003)}]{benabed03}
\bibinfo{author}{\bibfnamefont{K.}~\bibnamefont{{Benabed}}} \bibnamefont{and}
  \bibinfo{author}{\bibfnamefont{L.}~\bibnamefont{{Van Waerbeke}}},
  \bibinfo{journal}{ArXiv Astrophysics e-prints}  (\bibinfo{year}{2003}),
  \eprint{astro-ph/0306033}.

\bibitem[{\citenamefont{{Simon} et~al.}(2003)\citenamefont{{Simon}, {King}, and
  {Schneider}}}]{simon03}
\bibinfo{author}{\bibfnamefont{P.}~\bibnamefont{{Simon}}},
  \bibinfo{author}{\bibfnamefont{L.~J.} \bibnamefont{{King}}},
  \bibnamefont{and}
  \bibinfo{author}{\bibfnamefont{P.}~\bibnamefont{{Schneider}}},
  \bibinfo{journal}{ArXiv Astrophysics e-prints}  (\bibinfo{year}{2003}),
  \eprint{astro-ph/0309032}.

\bibitem[{\citenamefont{{Linder}}(2002)}]{linder02}
\bibinfo{author}{\bibfnamefont{E.~V.} \bibnamefont{{Linder}}},
  \bibinfo{journal}{ArXiv Astrophysics e-prints}  (\bibinfo{year}{2002}),
  \eprint{astro-ph/0200217}.

\bibitem[{\citenamefont{{Weller} and {Albrecht}}(2001)}]{weller01}
\bibinfo{author}{\bibfnamefont{J.}~\bibnamefont{{Weller}}} \bibnamefont{and}
  \bibinfo{author}{\bibfnamefont{A.}~\bibnamefont{{Albrecht}}},
  \bibinfo{journal}{Physical Review Letters} \textbf{\bibinfo{volume}{86}},
  \bibinfo{pages}{1939} (\bibinfo{year}{2001}).

\bibitem[{\citenamefont{{Huterer} and {Starkman}}(2003)}]{huterer03}
\bibinfo{author}{\bibfnamefont{D.}~\bibnamefont{{Huterer}}} \bibnamefont{and}
  \bibinfo{author}{\bibfnamefont{G.}~\bibnamefont{{Starkman}}},
  \bibinfo{journal}{Physical Review Letters} \textbf{\bibinfo{volume}{90}},
  \bibinfo{pages}{31301} (\bibinfo{year}{2003}).

\bibitem[{\citenamefont{Tyson et~al.}(2002)\citenamefont{Tyson, Wittman,
  Hennawi, and Spergel}}]{tyson02}
\bibinfo{author}{\bibfnamefont{J.~A.} \bibnamefont{Tyson}},
  \bibinfo{author}{\bibfnamefont{D.~M.} \bibnamefont{Wittman}},
  \bibinfo{author}{\bibfnamefont{J.~F.} \bibnamefont{Hennawi}},
  \bibnamefont{and} \bibinfo{author}{\bibfnamefont{D.~N.}
  \bibnamefont{Spergel}} (\bibinfo{year}{2002}), \eprint{astro-ph/0209632}.

\bibitem[{\citenamefont{{Jain} and {Taylor}}(2003)}]{jain03}
\bibinfo{author}{\bibfnamefont{B.}~\bibnamefont{{Jain}}} \bibnamefont{and}
  \bibinfo{author}{\bibfnamefont{A.}~\bibnamefont{{Taylor}}},
  \bibinfo{journal}{ArXiv Astrophysics e-prints}  (\bibinfo{year}{2003}),
  \eprint{astro-ph/0306046}.

\bibitem[{\citenamefont{{Bernstein} and {Jain}}(2003)}]{bernstein03}
\bibinfo{author}{\bibfnamefont{G.~M.} \bibnamefont{{Bernstein}}}
  \bibnamefont{and} \bibinfo{author}{\bibfnamefont{B.}~\bibnamefont{{Jain}}},
  \bibinfo{journal}{ArXiv Astrophysics e-prints}  (\bibinfo{year}{2003}),
  \eprint{astro-ph/0309332}.

\bibitem[{\citenamefont{{Hu}}(1999{\natexlab{a}})}]{hu99a}
\bibinfo{author}{\bibfnamefont{W.}~\bibnamefont{{Hu}}}, \bibinfo{journal}{ApJ}
  \textbf{\bibinfo{volume}{522}}, \bibinfo{pages}{L21}
  (\bibinfo{year}{1999}{\natexlab{a}}).

\bibitem[{\citenamefont{{Hu}}(1999{\natexlab{b}})}]{hu99b}
\bibinfo{author}{\bibfnamefont{W.}~\bibnamefont{{Hu}}},
  \textbf{\bibinfo{volume}{522}}, \bibinfo{pages}{L21}
  (\bibinfo{year}{1999}{\natexlab{b}}).

\bibitem[{\citenamefont{{Kaplinghat} et~al.}(2003)\citenamefont{{Kaplinghat},
  {Knox}, and {Song}}}]{kaplinghat03}
\bibinfo{author}{\bibfnamefont{M.}~\bibnamefont{{Kaplinghat}}},
  \bibinfo{author}{\bibfnamefont{L.}~\bibnamefont{{Knox}}}, \bibnamefont{and}
  \bibinfo{author}{\bibfnamefont{Y.}~\bibnamefont{{Song}}},
  \bibinfo{journal}{ArXiv Astrophysics e-prints} pp. \bibinfo{pages}{3344--+}
  (\bibinfo{year}{2003}).

\bibitem[{\citenamefont{{Beacom} and {Bell}}(2002)}]{beacom02}
\bibinfo{author}{\bibfnamefont{J.~F.} \bibnamefont{{Beacom}}} \bibnamefont{and}
  \bibinfo{author}{\bibfnamefont{N.~F.} \bibnamefont{{Bell}}},
  \bibinfo{journal}{\prd} \textbf{\bibinfo{volume}{65}},
  \bibinfo{pages}{113009} (\bibinfo{year}{2002}).

\bibitem[{\citenamefont{{Kaiser}}(1992)}]{kaiser92}
\bibinfo{author}{\bibfnamefont{N.}~\bibnamefont{{Kaiser}}},
  \bibinfo{journal}{\apj} \textbf{\bibinfo{volume}{388}}, \bibinfo{pages}{272}
  (\bibinfo{year}{1992}).

\bibitem[{\citenamefont{{Bartelmann} and {Schneider}}(2001)}]{bartelmann01}
\bibinfo{author}{\bibfnamefont{M.}~\bibnamefont{{Bartelmann}}}
  \bibnamefont{and}
  \bibinfo{author}{\bibfnamefont{P.}~\bibnamefont{{Schneider}}},
  \bibinfo{journal}{Phys. Rep.} \textbf{\bibinfo{volume}{340}},
  \bibinfo{pages}{291} (\bibinfo{year}{2001}).

\bibitem[{\citenamefont{{Kamionkowski}
  et~al.}(1997)\citenamefont{{Kamionkowski}, {Kosowsky}, and
  {Stebbins}}}]{kamionkowski97}
\bibinfo{author}{\bibfnamefont{M.}~\bibnamefont{{Kamionkowski}}},
  \bibinfo{author}{\bibfnamefont{A.}~\bibnamefont{{Kosowsky}}},
  \bibnamefont{and}
  \bibinfo{author}{\bibfnamefont{A.}~\bibnamefont{{Stebbins}}},
  \bibinfo{journal}{Phys. Rev. Lett.} \textbf{\bibinfo{volume}{78}},
  \bibinfo{pages}{2058} (\bibinfo{year}{1997}).

\bibitem[{\citenamefont{{Seljak} and {Zaldarriaga}}(1997)}]{seljak97}
\bibinfo{author}{\bibfnamefont{U.~.} \bibnamefont{{Seljak}}} \bibnamefont{and}
  \bibinfo{author}{\bibfnamefont{M.}~\bibnamefont{{Zaldarriaga}}},
  \bibinfo{journal}{Phys. Rev. Lett.} \textbf{\bibinfo{volume}{78}},
  \bibinfo{pages}{2054} (\bibinfo{year}{1997}).

\bibitem[{\citenamefont{{Peacock} and {Dodds}}(1994)}]{peacock94}
\bibinfo{author}{\bibfnamefont{J.~A.} \bibnamefont{{Peacock}}}
  \bibnamefont{and} \bibinfo{author}{\bibfnamefont{S.~J.}
  \bibnamefont{{Dodds}}}, \bibinfo{journal}{MNRAS}
  \textbf{\bibinfo{volume}{267}}, \bibinfo{pages}{1020} (\bibinfo{year}{1994}).

\bibitem[{\citenamefont{{Scoccimarro} et~al.}(1999)\citenamefont{{Scoccimarro},
  {Zaldarriaga}, and {Hui}}}]{scoccimarro99}
\bibinfo{author}{\bibfnamefont{R.}~\bibnamefont{{Scoccimarro}}},
  \bibinfo{author}{\bibfnamefont{M.}~\bibnamefont{{Zaldarriaga}}},
  \bibnamefont{and} \bibinfo{author}{\bibfnamefont{L.}~\bibnamefont{{Hui}}},
  \bibinfo{journal}{Astrophys. J.} \textbf{\bibinfo{volume}{527}},
  \bibinfo{pages}{1} (\bibinfo{year}{1999}).

\bibitem[{\citenamefont{{Hu} and {Okamoto}}(2002{\natexlab{b}})}]{hu02}
\bibinfo{author}{\bibfnamefont{W.}~\bibnamefont{{Hu}}} \bibnamefont{and}
  \bibinfo{author}{\bibfnamefont{T.}~\bibnamefont{{Okamoto}}},
  \textbf{\bibinfo{volume}{574}}, \bibinfo{pages}{566}
  (\bibinfo{year}{2002}{\natexlab{b}}).

\bibitem[{\citenamefont{{Hirata} and {Seljak}}(2003{\natexlab{a}})}]{hirata03b}
\bibinfo{author}{\bibfnamefont{C.~M.} \bibnamefont{{Hirata}}} \bibnamefont{and}
  \bibinfo{author}{\bibfnamefont{U.}~\bibnamefont{{Seljak}}},
  \bibinfo{journal}{ArXiv Astrophysics e-prints}
  (\bibinfo{year}{2003}{\natexlab{a}}), \eprint{astro-ph/0306354}.

\bibitem[{\citenamefont{{Coble} et~al.}(1997)\citenamefont{{Coble}, {Dodelson},
  and {Frieman}}}]{coble97}
\bibinfo{author}{\bibfnamefont{K.}~\bibnamefont{{Coble}}},
  \bibinfo{author}{\bibfnamefont{S.}~\bibnamefont{{Dodelson}}},
  \bibnamefont{and} \bibinfo{author}{\bibfnamefont{J.~A.}
  \bibnamefont{{Frieman}}}, \bibinfo{journal}{\prd}
  \textbf{\bibinfo{volume}{55}}, \bibinfo{pages}{1851} (\bibinfo{year}{1997}).

\bibitem[{\citenamefont{{Caldwell} et~al.}(1998)\citenamefont{{Caldwell},
  {Dave}, and {Steinhardt}}}]{caldwell98}
\bibinfo{author}{\bibfnamefont{R.~R.} \bibnamefont{{Caldwell}}},
  \bibinfo{author}{\bibfnamefont{R.}~\bibnamefont{{Dave}}}, \bibnamefont{and}
  \bibinfo{author}{\bibfnamefont{P.~J.} \bibnamefont{{Steinhardt}}},
  \bibinfo{journal}{Physical Review Letters} \textbf{\bibinfo{volume}{80}},
  \bibinfo{pages}{1582} (\bibinfo{year}{1998}).

\bibitem[{\citenamefont{{Ma} et~al.}(1999)\citenamefont{{Ma}, {Caldwell},
  {Bode}, and {Wang}}}]{ma99}
\bibinfo{author}{\bibfnamefont{C.}~\bibnamefont{{Ma}}},
  \bibinfo{author}{\bibfnamefont{R.~R.} \bibnamefont{{Caldwell}}},
  \bibinfo{author}{\bibfnamefont{P.}~\bibnamefont{{Bode}}}, \bibnamefont{and}
  \bibinfo{author}{\bibfnamefont{L.}~\bibnamefont{{Wang}}},
  \bibinfo{journal}{ApJ} \textbf{\bibinfo{volume}{521}}, \bibinfo{pages}{L1}
  (\bibinfo{year}{1999}).

\bibitem[{\citenamefont{{Contaldi} et~al.}(2003)\citenamefont{{Contaldi},
  {Hoekstra}, and {Lewis}}}]{contaldi03}
\bibinfo{author}{\bibfnamefont{C.~R.} \bibnamefont{{Contaldi}}},
  \bibinfo{author}{\bibfnamefont{H.}~\bibnamefont{{Hoekstra}}},
  \bibnamefont{and} \bibinfo{author}{\bibfnamefont{A.}~\bibnamefont{{Lewis}}},
  \bibinfo{journal}{Physical Review Letters} \textbf{\bibinfo{volume}{90}},
  \bibinfo{pages}{221303} (\bibinfo{year}{2003}).

\bibitem[{\citenamefont{{Hoekstra} et~al.}(2002)\citenamefont{{Hoekstra},
  {Yee}, and {Gladders}}}]{hoekstra02}
\bibinfo{author}{\bibfnamefont{H.}~\bibnamefont{{Hoekstra}}},
  \bibinfo{author}{\bibfnamefont{H.~K.~C.} \bibnamefont{{Yee}}},
  \bibnamefont{and} \bibinfo{author}{\bibfnamefont{M.~D.}
  \bibnamefont{{Gladders}}}, \bibinfo{journal}{\apj}
  \textbf{\bibinfo{volume}{577}}, \bibinfo{pages}{595} (\bibinfo{year}{2002}).

\bibitem[{\citenamefont{{Nagashima} et~al.}(2002)\citenamefont{{Nagashima},
  {Yoshii}, {Totani}, and {Gouda}}}]{nagashima02}
\bibinfo{author}{\bibfnamefont{M.}~\bibnamefont{{Nagashima}}},
  \bibinfo{author}{\bibfnamefont{Y.}~\bibnamefont{{Yoshii}}},
  \bibinfo{author}{\bibfnamefont{T.}~\bibnamefont{{Totani}}}, \bibnamefont{and}
  \bibinfo{author}{\bibfnamefont{N.}~\bibnamefont{{Gouda}}},
  \bibinfo{journal}{ArXiv Astrophysics e-prints} pp. \bibinfo{pages}{7483--+}
  (\bibinfo{year}{2002}).

\bibitem[{\citenamefont{{Massey} et~al.}(2003)\citenamefont{{Massey}, {Rhodes},
  {Refregier}, {Albert}, {Bacon}, {Bernstein}, {Ellis}, {Jain}, {McKay},
  {Perlmutter} et~al.}}]{massey03}
\bibinfo{author}{\bibfnamefont{R.}~\bibnamefont{{Massey}}},
  \bibinfo{author}{\bibfnamefont{J.}~\bibnamefont{{Rhodes}}},
  \bibinfo{author}{\bibfnamefont{A.}~\bibnamefont{{Refregier}}},
  \bibinfo{author}{\bibfnamefont{J.}~\bibnamefont{{Albert}}},
  \bibinfo{author}{\bibfnamefont{D.}~\bibnamefont{{Bacon}}},
  \bibinfo{author}{\bibfnamefont{G.}~\bibnamefont{{Bernstein}}},
  \bibinfo{author}{\bibfnamefont{R.}~\bibnamefont{{Ellis}}},
  \bibinfo{author}{\bibfnamefont{B.}~\bibnamefont{{Jain}}},
  \bibinfo{author}{\bibfnamefont{T.}~\bibnamefont{{McKay}}},
  \bibinfo{author}{\bibfnamefont{S.}~\bibnamefont{{Perlmutter}}},
  \bibnamefont{et~al.}, \bibinfo{journal}{ArXiv Astrophysics e-prints}
  (\bibinfo{year}{2003}), \eprint{astro-ph/0304418}.

\bibitem[{\citenamefont{{Kaiser}}(1998)}]{kaiser98}
\bibinfo{author}{\bibfnamefont{N.}~\bibnamefont{{Kaiser}}},
  \bibinfo{journal}{\apj} \textbf{\bibinfo{volume}{498}}, \bibinfo{pages}{26+}
  (\bibinfo{year}{1998}).

\bibitem[{\citenamefont{{Jarvis} et~al.}(2003)\citenamefont{{Jarvis},
  {Bernstein}, {Fischer}, {Smith}, {Jain}, {Tyson}, and {Wittman}}}]{jarvis03}
\bibinfo{author}{\bibfnamefont{M.}~\bibnamefont{{Jarvis}}},
  \bibinfo{author}{\bibfnamefont{G.~M.} \bibnamefont{{Bernstein}}},
  \bibinfo{author}{\bibfnamefont{P.}~\bibnamefont{{Fischer}}},
  \bibinfo{author}{\bibfnamefont{D.}~\bibnamefont{{Smith}}},
  \bibinfo{author}{\bibfnamefont{B.}~\bibnamefont{{Jain}}},
  \bibinfo{author}{\bibfnamefont{J.~A.} \bibnamefont{{Tyson}}},
  \bibnamefont{and}
  \bibinfo{author}{\bibfnamefont{D.}~\bibnamefont{{Wittman}}},
  \bibinfo{journal}{Astron. J.} \textbf{\bibinfo{volume}{125}},
  \bibinfo{pages}{1014} (\bibinfo{year}{2003}).

\bibitem[{\citenamefont{{Tauber}}(2001)}]{planck}
\bibinfo{author}{\bibfnamefont{J.~A.} \bibnamefont{{Tauber}}}, in
  \emph{\bibinfo{booktitle}{IAU Symposium}} (\bibinfo{year}{2001}), pp.
  \bibinfo{pages}{493--+}.

\bibitem[{\citenamefont{{Knox}}(1999)}]{knox99}
\bibinfo{author}{\bibfnamefont{L.}~\bibnamefont{{Knox}}},
  \bibinfo{journal}{MNRAS} \textbf{\bibinfo{volume}{307}}, \bibinfo{pages}{977}
  (\bibinfo{year}{1999}).

\bibitem[{\citenamefont{{Tegmark} et~al.}(2000)\citenamefont{{Tegmark},
  {Eisenstein}, {Hu}, and {de Oliveira-Costa}}}]{tegmark00b}
\bibinfo{author}{\bibfnamefont{M.}~\bibnamefont{{Tegmark}}},
  \bibinfo{author}{\bibfnamefont{D.~J.} \bibnamefont{{Eisenstein}}},
  \bibinfo{author}{\bibfnamefont{W.}~\bibnamefont{{Hu}}}, \bibnamefont{and}
  \bibinfo{author}{\bibfnamefont{A.}~\bibnamefont{{de Oliveira-Costa}}},
  \bibinfo{journal}{\apj} \textbf{\bibinfo{volume}{530}}, \bibinfo{pages}{133}
  (\bibinfo{year}{2000}).

\bibitem[{\citenamefont{{Bouchet} and {Gispert}}(1999)}]{bouchet99}
\bibinfo{author}{\bibfnamefont{F.~R.} \bibnamefont{{Bouchet}}}
  \bibnamefont{and}
  \bibinfo{author}{\bibfnamefont{R.}~\bibnamefont{{Gispert}}},
  \bibinfo{journal}{New Astronomy} \textbf{\bibinfo{volume}{4}},
  \bibinfo{pages}{443} (\bibinfo{year}{1999}).

\bibitem[{\citenamefont{{Jungman} et~al.}(1996)\citenamefont{{Jungman},
  {Kamionkowski}, {Kosowsky}, and {Spergel}}}]{jungman96}
\bibinfo{author}{\bibfnamefont{G.}~\bibnamefont{{Jungman}}},
  \bibinfo{author}{\bibfnamefont{M.}~\bibnamefont{{Kamionkowski}}},
  \bibinfo{author}{\bibfnamefont{A.}~\bibnamefont{{Kosowsky}}},
  \bibnamefont{and} \bibinfo{author}{\bibfnamefont{D.~N.}
  \bibnamefont{{Spergel}}}, \bibinfo{journal}{\prd}
  \textbf{\bibinfo{volume}{54}}, \bibinfo{pages}{1332} (\bibinfo{year}{1996}).

\bibitem[{\citenamefont{{Eisenstein} et~al.}(1999)\citenamefont{{Eisenstein},
  {Hu}, and {Tegmark}}}]{eisenstein99}
\bibinfo{author}{\bibfnamefont{D.~J.} \bibnamefont{{Eisenstein}}},
  \bibinfo{author}{\bibfnamefont{W.}~\bibnamefont{{Hu}}}, \bibnamefont{and}
  \bibinfo{author}{\bibfnamefont{M.}~\bibnamefont{{Tegmark}}},
  \bibinfo{journal}{\apj} \textbf{\bibinfo{volume}{518}}, \bibinfo{pages}{2}
  (\bibinfo{year}{1999}).

\bibitem[{\citenamefont{{Efstathiou} and {Bond}}(1999)}]{efstathiou99}
\bibinfo{author}{\bibfnamefont{G.}~\bibnamefont{{Efstathiou}}}
  \bibnamefont{and} \bibinfo{author}{\bibfnamefont{J.~R.}
  \bibnamefont{{Bond}}}, \textbf{\bibinfo{volume}{304}}, \bibinfo{pages}{75}
  (\bibinfo{year}{1999}).

\bibitem[{\citenamefont{{Hu} et~al.}(2001)\citenamefont{{Hu}, {Fukugita},
  {Zaldarriaga}, and {Tegmark}}}]{hu01a}
\bibinfo{author}{\bibfnamefont{W.}~\bibnamefont{{Hu}}},
  \bibinfo{author}{\bibfnamefont{M.}~\bibnamefont{{Fukugita}}},
  \bibinfo{author}{\bibfnamefont{M.}~\bibnamefont{{Zaldarriaga}}},
  \bibnamefont{and}
  \bibinfo{author}{\bibfnamefont{M.}~\bibnamefont{{Tegmark}}},
  \bibinfo{journal}{Astrophys. J.} \textbf{\bibinfo{volume}{549}},
  \bibinfo{pages}{669} (\bibinfo{year}{2001}).

\bibitem[{\citenamefont{{Kosowsky} et~al.}(2002)\citenamefont{{Kosowsky},
  {Milosavljevic}, and {Jimenez}}}]{kosowsky02}
\bibinfo{author}{\bibfnamefont{A.}~\bibnamefont{{Kosowsky}}},
  \bibinfo{author}{\bibfnamefont{M.}~\bibnamefont{{Milosavljevic}}},
  \bibnamefont{and}
  \bibinfo{author}{\bibfnamefont{R.}~\bibnamefont{{Jimenez}}},
  \bibinfo{journal}{\prd} \textbf{\bibinfo{volume}{66}}, \bibinfo{pages}{63007}
  (\bibinfo{year}{2002}).

\bibitem[{\citenamefont{{Zaldarriaga} et~al.}(1997)\citenamefont{{Zaldarriaga},
  {Spergel}, and {Seljak}}}]{zaldarriaga97}
\bibinfo{author}{\bibfnamefont{M.}~\bibnamefont{{Zaldarriaga}}},
  \bibinfo{author}{\bibfnamefont{D.~N.} \bibnamefont{{Spergel}}},
  \bibnamefont{and} \bibinfo{author}{\bibfnamefont{U.}~\bibnamefont{{Seljak}}},
  \bibinfo{journal}{\apj} \textbf{\bibinfo{volume}{488}}, \bibinfo{pages}{1+}
  (\bibinfo{year}{1997}).

\bibitem[{\citenamefont{{Knox} et~al.}(2001)\citenamefont{{Knox}, {Cooray},
  {Eisenstein}, and {Haiman}}}]{knox01a}
\bibinfo{author}{\bibfnamefont{L.}~\bibnamefont{{Knox}}},
  \bibinfo{author}{\bibfnamefont{A.}~\bibnamefont{{Cooray}}},
  \bibinfo{author}{\bibfnamefont{D.}~\bibnamefont{{Eisenstein}}},
  \bibnamefont{and} \bibinfo{author}{\bibfnamefont{Z.}~\bibnamefont{{Haiman}}},
  \bibinfo{journal}{\apj} \textbf{\bibinfo{volume}{550}}, \bibinfo{pages}{7}
  (\bibinfo{year}{2001}).

\bibitem[{\citenamefont{Bond and Szalay}(1983)}]{bond83}
\bibinfo{author}{\bibfnamefont{J.~R.} \bibnamefont{Bond}} \bibnamefont{and}
  \bibinfo{author}{\bibfnamefont{A.~S.} \bibnamefont{Szalay}},
  \bibinfo{journal}{Astrophys. J.} \textbf{\bibinfo{volume}{274}},
  \bibinfo{pages}{443} (\bibinfo{year}{1983}).

\bibitem[{\citenamefont{{Ma}}(1996)}]{ma96}
\bibinfo{author}{\bibfnamefont{C.}~\bibnamefont{{Ma}}}, \bibinfo{journal}{\apj}
  \textbf{\bibinfo{volume}{471}}, \bibinfo{pages}{13} (\bibinfo{year}{1996}).

\bibitem[{\citenamefont{{Hu} and {Eisenstein}}(1998)}]{hu98}
\bibinfo{author}{\bibfnamefont{W.}~\bibnamefont{{Hu}}} \bibnamefont{and}
  \bibinfo{author}{\bibfnamefont{D.~J.} \bibnamefont{{Eisenstein}}},
  \bibinfo{journal}{\apj} \textbf{\bibinfo{volume}{498}}, \bibinfo{pages}{497}
  (\bibinfo{year}{1998}).

\bibitem[{\citenamefont{{Hu} et~al.}(1998)\citenamefont{{Hu}, {Eisenstein}, and
  {Tegmark}}}]{hu98a}
\bibinfo{author}{\bibfnamefont{W.}~\bibnamefont{{Hu}}},
  \bibinfo{author}{\bibfnamefont{D.~J.} \bibnamefont{{Eisenstein}}},
  \bibnamefont{and}
  \bibinfo{author}{\bibfnamefont{M.}~\bibnamefont{{Tegmark}}},
  \bibinfo{journal}{Phys. Rev. Lett.} \textbf{\bibinfo{volume}{80}},
  \bibinfo{pages}{5255} (\bibinfo{year}{1998}).

\bibitem[{\citenamefont{Spergel et~al.}(2003)}]{spergel03}
\bibinfo{author}{\bibfnamefont{D.~N.} \bibnamefont{Spergel}}
  \bibnamefont{et~al.} (\bibinfo{year}{2003}),
  \bibinfo{note}{astro-ph/0302209}.

\bibitem[{\citenamefont{{Seljak} et~al.}(2003)\citenamefont{{Seljak},
  {McDonald}, and {Makarov}}}]{seljak03}
\bibinfo{author}{\bibfnamefont{U.}~\bibnamefont{{Seljak}}},
  \bibinfo{author}{\bibfnamefont{P.}~\bibnamefont{{McDonald}}},
  \bibnamefont{and}
  \bibinfo{author}{\bibfnamefont{A.}~\bibnamefont{{Makarov}}},
  \bibinfo{journal}{ArXiv Astrophysics e-prints} pp. \bibinfo{pages}{2571--+}
  (\bibinfo{year}{2003}).

\bibitem[{\citenamefont{{Zaldarriaga} et~al.}(2001)\citenamefont{{Zaldarriaga},
  {Hui}, and {Tegmark}}}]{zaldarriaga01}
\bibinfo{author}{\bibfnamefont{M.}~\bibnamefont{{Zaldarriaga}}},
  \bibinfo{author}{\bibfnamefont{L.}~\bibnamefont{{Hui}}}, \bibnamefont{and}
  \bibinfo{author}{\bibfnamefont{M.}~\bibnamefont{{Tegmark}}},
  \bibinfo{journal}{Astrophys. J.} \textbf{\bibinfo{volume}{557}},
  \bibinfo{pages}{519} (\bibinfo{year}{2001}).

\bibitem[{\citenamefont{{Gold} and {Albrecht}}(2003)}]{gold03}
\bibinfo{author}{\bibfnamefont{B.}~\bibnamefont{{Gold}}} \bibnamefont{and}
  \bibinfo{author}{\bibfnamefont{A.}~\bibnamefont{{Albrecht}}},
  \bibinfo{journal}{ArXiv Astrophysics e-prints}  (\bibinfo{year}{2003}),
  \eprint{astro-ph/0301050}.

\bibitem[{\citenamefont{{Ishak} et~al.}(2003)\citenamefont{{Ishak}, {Hirata},
  {McDonald}, and {Seljak}}}]{ishak03}
\bibinfo{author}{\bibfnamefont{M.}~\bibnamefont{{Ishak}}},
  \bibinfo{author}{\bibfnamefont{C.~M.} \bibnamefont{{Hirata}}},
  \bibinfo{author}{\bibfnamefont{P.}~\bibnamefont{{McDonald}}},
  \bibnamefont{and} \bibinfo{author}{\bibfnamefont{U.}~\bibnamefont{{Seljak}}},
  \bibinfo{journal}{ArXiv Astrophysics e-prints}  (\bibinfo{year}{2003}),
  \eprint{astro-ph/0308446}.

\bibitem[{\citenamefont{{Knox}}(2003)}]{knox03}
\bibinfo{author}{\bibfnamefont{L.}~\bibnamefont{{Knox}}},
  \bibinfo{journal}{ArXiv Astrophysics e-prints} pp. \bibinfo{pages}{4370--+}
  (\bibinfo{year}{2003}).

\bibitem[{\citenamefont{{Takada} and {Jain}}(2003{\natexlab{b}})}]{takada03a}
\bibinfo{author}{\bibfnamefont{M.}~\bibnamefont{{Takada}}} \bibnamefont{and}
  \bibinfo{author}{\bibfnamefont{B.}~\bibnamefont{{Jain}}},
  \bibinfo{journal}{ArXiv Astrophysics e-prints}
  (\bibinfo{year}{2003}{\natexlab{b}}), \eprint{astro-ph/0310125}.

\bibitem[{\citenamefont{{Albrecht} and {Skordis}}(2000)}]{albrecht00}
\bibinfo{author}{\bibfnamefont{A.}~\bibnamefont{{Albrecht}}} \bibnamefont{and}
  \bibinfo{author}{\bibfnamefont{C.}~\bibnamefont{{Skordis}}},
  \bibinfo{journal}{Physical Review Letters} \textbf{\bibinfo{volume}{84}},
  \bibinfo{pages}{2076} (\bibinfo{year}{2000}).

\bibitem[{\citenamefont{{Armendariz-Picon}
  et~al.}(2000)\citenamefont{{Armendariz-Picon}, {Mukhanov}, and
  {Steinhardt}}}]{armendariz00}
\bibinfo{author}{\bibfnamefont{C.}~\bibnamefont{{Armendariz-Picon}}},
  \bibinfo{author}{\bibfnamefont{V.}~\bibnamefont{{Mukhanov}}},
  \bibnamefont{and} \bibinfo{author}{\bibfnamefont{P.~J.}
  \bibnamefont{{Steinhardt}}}, \bibinfo{journal}{Physical Review Letters}
  \textbf{\bibinfo{volume}{85}}, \bibinfo{pages}{4438} (\bibinfo{year}{2000}).

\bibitem[{\citenamefont{{Farrar} and {Peebles}}(2003)}]{farrar03}
\bibinfo{author}{\bibfnamefont{G.~R.} \bibnamefont{{Farrar}}} \bibnamefont{and}
  \bibinfo{author}{\bibfnamefont{P.~J.~E.} \bibnamefont{{Peebles}}},
  \bibinfo{journal}{ArXiv Astrophysics e-prints}  (\bibinfo{year}{2003}),
  \eprint{astro-ph/0307316}.

\bibitem[{\citenamefont{{Hirata} and {Seljak}}(2003{\natexlab{b}})}]{hirata03c}
\bibinfo{author}{\bibfnamefont{C.}~\bibnamefont{{Hirata}}} \bibnamefont{and}
  \bibinfo{author}{\bibfnamefont{U.}~\bibnamefont{{Seljak}}},
  \textbf{\bibinfo{volume}{343}}, \bibinfo{pages}{459}
  (\bibinfo{year}{2003}{\natexlab{b}}).

\bibitem[{\citenamefont{{Heavens} et~al.}(2000)\citenamefont{{Heavens},
  {Refregier}, and {Heymans}}}]{heavens00}
\bibinfo{author}{\bibfnamefont{A.}~\bibnamefont{{Heavens}}},
  \bibinfo{author}{\bibfnamefont{A.}~\bibnamefont{{Refregier}}},
  \bibnamefont{and}
  \bibinfo{author}{\bibfnamefont{C.}~\bibnamefont{{Heymans}}},
  \textbf{\bibinfo{volume}{319}}, \bibinfo{pages}{649} (\bibinfo{year}{2000}).

\bibitem[{\citenamefont{{Catelan} et~al.}(2001)\citenamefont{{Catelan},
  {Kamionkowski}, and {Blandford}}}]{catelan01}
\bibinfo{author}{\bibfnamefont{P.}~\bibnamefont{{Catelan}}},
  \bibinfo{author}{\bibfnamefont{M.}~\bibnamefont{{Kamionkowski}}},
  \bibnamefont{and} \bibinfo{author}{\bibfnamefont{R.~D.}
  \bibnamefont{{Blandford}}}, \textbf{\bibinfo{volume}{320}},
  \bibinfo{pages}{L7} (\bibinfo{year}{2001}).

\bibitem[{\citenamefont{{Crittenden} et~al.}(2001)\citenamefont{{Crittenden},
  {Natarajan}, {Pen}, and {Theuns}}}]{crittenden01}
\bibinfo{author}{\bibfnamefont{R.~G.} \bibnamefont{{Crittenden}}},
  \bibinfo{author}{\bibfnamefont{P.}~\bibnamefont{{Natarajan}}},
  \bibinfo{author}{\bibfnamefont{U.}~\bibnamefont{{Pen}}}, \bibnamefont{and}
  \bibinfo{author}{\bibfnamefont{T.}~\bibnamefont{{Theuns}}},
  \bibinfo{journal}{\apj} \textbf{\bibinfo{volume}{559}}, \bibinfo{pages}{552}
  (\bibinfo{year}{2001}).

\bibitem[{\citenamefont{{Pen} et~al.}(2000)\citenamefont{{Pen}, {Lee}, and
  {Seljak}}}]{pen00}
\bibinfo{author}{\bibfnamefont{U.}~\bibnamefont{{Pen}}},
  \bibinfo{author}{\bibfnamefont{J.}~\bibnamefont{{Lee}}}, \bibnamefont{and}
  \bibinfo{author}{\bibfnamefont{U.}~\bibnamefont{{Seljak}}},
  \textbf{\bibinfo{volume}{543}}, \bibinfo{pages}{L107} (\bibinfo{year}{2000}).

\bibitem[{\citenamefont{{Mackey} et~al.}(2002)\citenamefont{{Mackey}, {White},
  and {Kamionkowski}}}]{mackey02}
\bibinfo{author}{\bibfnamefont{J.}~\bibnamefont{{Mackey}}},
  \bibinfo{author}{\bibfnamefont{M.}~\bibnamefont{{White}}}, \bibnamefont{and}
  \bibinfo{author}{\bibfnamefont{M.}~\bibnamefont{{Kamionkowski}}},
  \textbf{\bibinfo{volume}{332}}, \bibinfo{pages}{788} (\bibinfo{year}{2002}).

\bibitem[{\citenamefont{{Jing}}(2002)}]{jing02}
\bibinfo{author}{\bibfnamefont{Y.~P.} \bibnamefont{{Jing}}},
  \textbf{\bibinfo{volume}{335}}, \bibinfo{pages}{L89} (\bibinfo{year}{2002}).

\bibitem[{\citenamefont{{Takada} and {White}}(2003)}]{takada03c}
\bibinfo{author}{\bibfnamefont{M.}~\bibnamefont{{Takada}}} \bibnamefont{and}
  \bibinfo{author}{\bibfnamefont{M.}~\bibnamefont{{White}}},
  \bibinfo{journal}{ArXiv Astrophysics e-prints}  (\bibinfo{year}{2003}),
  \eprint{astro-ph/0311104}.

\end{thebibliography}

\end{document}